\documentstyle[12pt,epsfig,amssymb,enumerate,xcolor]{article}

\newcommand{\eV}{\textrm{eV}}
\newcommand{\fns}{\footnotesize}

\begin{document}

\vspace{0.2cm}

\begin{center}
{\large \bf Hybrid Textures of Majorana Neutrino Mass Matrix and
\\ Current Experimental Tests}
\end{center}

\vspace{0.2cm}
\begin{center}
{\bf Ji-Yuan Liu} \footnote{E-mail: liujy@tjut.edu.cn} \\
{\sl College of Science, Tianjin University of Technology, Tianjin
300384, China} \\
\vspace{0.7cm}
{\bf Shun Zhou} \footnote{E-mail: shunzhou@kth.se} \\
{\sl Department of Theoretical Physics, School of Engineering
Sciences, \\ KTH Royal Institute of Technology, 106 91 Stockholm,
Sweden}
\end{center}

\vspace{2.0cm}

\begin{abstract}
Motivated by recent measurements of a relatively large
$\theta^{}_{13}$ in the Daya Bay and RENO reactor neutrino
experiments, we carry out a systematic analysis of the hybrid
textures of Majorana neutrino mass matrix $M^{}_\nu$, which contain
one texture zero and two equal nonzero matrix elements. We show that
three neutrino masses $(m^{}_1, m^{}_2, m^{}_3)$ and three leptonic
CP-violating phases $(\delta, \rho, \sigma)$ can fully be determined
from two neutrino mass-squared differences $(\delta m^2, \Delta
m^2)$ and three flavor mixing angles $(\theta^{}_{12},
\theta^{}_{23}, \theta^{}_{13})$. Out of sixty logically possible
patterns of $M^{}_\nu$, thirty-nine are found to be compatible with
current experimental data at the $3\sigma$ level. We demonstrate
that the texture zero of $M^{}_\nu$ is stable against one-loop
quantum corrections, while the equality between two independent
elements not. Phenomenological implications of $M^{}_\nu$ for the
neutrinoless double-beta decay and leptonic CP violation are
discussed, and a realization of the texture zero and equality by
means of discrete flavor symmetries is illustrated.
\end{abstract}
\begin{center}
\end{center}

\newpage

\section{Introduction}

Recent years have seen great progress in neutrino physics \cite{XZ}.
Thanks to a number of elegant solar, atmospheric, accelerator and
reactor neutrino oscillation experiments \cite{PDG}, two neutrino
mixing angles are found to be quite large (i.e., $\theta^{}_{12}
\approx 34^\circ$ and $\theta^{}_{23} \approx 40^\circ$), while two
independent neutrino mass-squared differences $\delta m^2 \equiv
m^2_2 - m^2_1$ and $\Delta m^2 \equiv m^2_3 - (m^2_1 + m^2_2)/2$ are
measured with a good degree of accuracy (i.e., $\delta m^2 \approx
7.5\times 10^{-5}~{\rm eV}^2$ and $|\Delta m^2| \approx 2.5\times
10^{-3}~{\rm eV}^2$). The latest results from the Daya Bay
\cite{Daya} and RENO \cite{Reno} experiments reveal that
$\theta^{}_{13} \approx 9^\circ$ is relatively large, which is very
crucial to determine the neutrino mass hierarchy (i.e., the sign of
$\Delta m^2$) and to discover the leptonic CP violation (i.e., the
Dirac CP-violating phase $\delta$) in the future long-baseline
neutrino oscillation experiments. However, the absolute scale of
neutrino masses and whether neutrinos are Dirac or Majorana
particles are still unknown.

On the theoretical side, a satisfactory description of tiny neutrino
masses and leptonic mixing pattern is still lacking. Although the
seesaw mechanisms can be responsible for the generation of light
neutrino masses \cite{SS1,SS2,SS3}, they leave the lepton flavor
structure intact. In fact, it was shown one decade ago that the
large leptonic mixing can be achieved by taking two independent
elements of Majorana neutrino mass matrix $M^{}_\nu$ to be zero, in
the flavor basis where the charged-lepton mass matrix $M^{}_l$ is
diagonal \cite{FGM,xing1,xing2,Guo}. Recently, several authors have
demonstrated that these seven two-zero textures of $M^{}_\nu$ still
survive the current neutrino oscillation data \cite{FXZ,Ludl}.
Furthermore, it has been pointed out that those texture zeros can be
realized by implementing the $Z^{}_n$ flavor symmetry in the type-II
seesaw model, where the Higgs triplets are introduced to account for
tiny Majorana neutrino masses \cite{FXZ}. Apart from texture zeros
in the neutrino mass matrix, possible correlations between two
matrix elements of $M^{}_\nu$ have recently been investigated
\cite{Grimus}.

In the present paper, we perform a systematic study of $M^{}_\nu$
with one texture zero and two equal nonzero elements, which has been
termed as ``hybrid texture" in the literature
\cite{Kaneko,Dev,hybrids}. The motivation for such an investigation
is three-fold. First, from the phenomenological point of view,
either one texture zero or an equality between two independent
entries in $M^{}_\nu$ imposes one constraint condition and thus
reduces the number of real free model parameters by two. Hence the
hybrid textures are as predictive as the well-known two-zero
textures, and deserve a detailed analysis. See, e.g., Refs.
\cite{Kaneko} and \cite{Dev}, for previous studies of hybrid
textures. The textures with two equalities and other
phenomenological assumptions have also been considered
\cite{Dev2013,Frigerio}. Second, it has been proved that a texture
zero in any position in $M^{}_\nu$ can be realized by using Abelian
flavor symmetries $Z^{}_n$ \cite{Grimus1,Grimus2} or $U(1)$
\cite{Heeck}. However, the equality between two nonzero matrix
elements should come from a non-Abelian flavor symmetry. Third, now
that a good knowledge about three neutrino mixing angles and two
neutrino mass-squared differences has been obtained, it is timely to
reexamine the possible structure of $M^{}_\nu$ and explore the
underlying symmetry in the lepton sector. Taking into account
current neutrino oscillation data at the $3\sigma$ level, we have
found that thirty-nine out of sixty logically possible hybrid
textures of $M^{}_\nu$ are viable. If the $1\sigma$ ranges of
neutrino mixing parameters are considered, only thirteen hybrid
textures can survive.

The remaining part of our paper is organized as follows. In
section~2, we introduce the hybrid textures and present some useful
analytical formulas. Then, the stability of texture zeros and
equalities in $M^{}_\nu$ against quantum corrections is briefly
discussed by using the one-loop renormalization group equation.
Section 3 is devoted to the analytical analysis of six viable
patterns, which serve as typical examples of hybrid textures. We
show that three neutrino masses $(m^{}_1, m^{}_2, m^{}_3)$ and three
CP-violating phases $(\delta, \rho, \sigma)$ can fully be determined
from neutrino mixing angles $(\theta^{}_{12}, \theta^{}_{23},
\theta^{}_{13})$ and neutrino mass-squared differences $(\delta m^2,
\Delta m^2)$. In section~4, a thorough numerical analysis has been
performed, and the allowed parameter space of the chosen six viable
patterns is given. To illustrate how to realize a hybrid texture, we
give a concrete example in section~5, where a type-II seesaw model
with an $S^{}_3\otimes Z^{}_3$ flavor symmetry is considered.
Finally, we summarize our main conclusions in section~6.

\section{Hybrid Textures}

\subsection{Classification}

At low energies, lepton mass spectra and flavor mixing are
determined by the charged-lepton mass matrix $M^{}_l$ and the
effective neutrino mass matrix $M^{}_\nu$. We assume massive
neutrinos to be Majorana particles, as in various seesaw models
\cite{SS1,SS2,SS3}, so $M^{}_\nu$ is in general a $3\times 3$
symmetric complex matrix. If one of six independent matrix elements
of $M^{}_\nu$ is taken to be zero and two of the rest are equal, we
finally arrive at ${\rm C}^1_6 \cdot {\rm C}^2_5 = 60$ logically
possible textures.

We enumerate all the 39 hybrid textures, which are compatible with
current neutrino oscillation data at the $3\sigma$ level and can be
classified into six categories:
\begin{eqnarray}
&& {\bf A}^{}_1:
  \left(\matrix{
    0 & \times & \times\cr
    \times & \triangle & \triangle\cr
    \times & \triangle & \times}
  \right),~~~~
  {\bf A}^{}_2:
  \left(\matrix{
    0 & \times & \times\cr
    \times & \triangle & \times\cr
    \times & \times & \triangle}
  \right),~~~~
  {\bf A}^{}_3:
  \left(\matrix{
    0 & \times & \times\cr
    \times & \times & \triangle \cr
    \times & \triangle & \triangle}
    \right) ;
%     (1)
\end{eqnarray}

\begin{eqnarray}
&& {\bf B}^{}_1: \left(\matrix{
    \triangle & 0 & \triangle \cr
    0 & \times & \times \cr
    \triangle & \times & \times}
    \right),~~~~
{\bf B}^{}_2:\left(\matrix{
    \times & 0 & \triangle \cr
    0 & \triangle & \times \cr
    \triangle & \times & \times}
  \right), ~~~~
{\bf B}^{}_3:\left(\matrix{
    \times & 0 & \triangle \cr
    0 & \times & \times \cr
    \triangle & \times & \triangle}
  \right),\nonumber \\
&&    {\bf B}^{}_4:\left(\matrix{
    \times & 0 & \times \cr
    0 & \triangle & \triangle \cr
    \times & \triangle & \times}
  \right),~~~~
  {\bf B}^{}_5:\left(\matrix{
    \times & 0 & \times \cr
    0 & \times & \triangle \cr
    \times & \triangle & \triangle}
  \right);
  %     (2)
\end{eqnarray}

\begin{eqnarray}
&& {\bf C}^{}_1:\left(\matrix{
    \triangle & \triangle & 0 \cr
    \triangle & \times & \times \cr
    0 & \times & \times}
  \right),~~~~
{\bf C}^{}_2:\left(\matrix{
    \times & \triangle & 0 \cr
    \triangle & \triangle & \times \cr
    0 & \times & \times}
  \right), ~~~~
{\bf C}^{}_3:\left(\matrix{
    \times & \triangle & 0 \cr
    \triangle & \times & \times \cr
    0 & \times & \triangle}
    \right),\nonumber \\
&& {\bf C}^{}_4:\left(\matrix{
    \times & \times & 0 \cr
    \times & \triangle & \triangle \cr
    0 & \triangle & \times}
     \right),~~~~
{\bf C}^{}_5:\left(\matrix{
    \times & \times & 0 \cr
    \times & \times & \triangle \cr
    0 & \triangle & \triangle}
     \right);
%     (3)
\end{eqnarray}

\begin{eqnarray}
&& {\bf D}^{}_1:\left(\matrix{
    \triangle & \triangle & \times \cr
    \triangle & \times & 0 \cr
    \times & 0 & \times}
  \right),~~~~
 {\bf D}^{}_2:\left(\matrix{
    \triangle & \times & \triangle \cr
    \times & \times & 0 \cr
    \triangle & 0 & \times}
  \right),~~~~
 {\bf D}^{}_3:\left(\matrix{
    \times & \triangle & \times \cr
    \triangle & \triangle & 0 \cr
    \times & 0 & \times}
  \right), \nonumber \\
&& {\bf D}^{}_4:\left(\matrix{
    \times & \triangle & \times \cr
    \triangle & \times & 0 \cr
    \times & 0 & \triangle}
  \right),~~~~
  {\bf D}^{}_5:\left(\matrix{
    \times & \times & \triangle \cr
    \times & \triangle & 0 \cr
    \triangle & 0 & \times}
  \right),~~~~
  {\bf D}^{}_6:\left(\matrix{
    \times & \times & \triangle \cr
    \times & \times & 0 \cr
    \triangle & 0 & \triangle}
  \right);
%     (4)
\end{eqnarray}
\vspace{0.2cm}
\begin{eqnarray}
&& {\bf E}^{}_1:\left(\matrix{
    \triangle & \triangle & \times \cr
    \triangle & 0 & \times \cr
    \times & \times & \times}
  \right),~
  {\bf E}^{}_2:\left(\matrix{
    \triangle & \times & \triangle \cr
    \times & 0 & \times \cr
    \triangle & \times & \times}
  \right),~
  {\bf E}^{}_3:\left(\matrix{
    \triangle & \times & \times \cr
    \times & 0 & \triangle \cr
    \times & \triangle & \times}
  \right),~
  {\bf E}^{}_4:\left(\matrix{
    \triangle & \times & \times \cr
    \times & 0 & \times \cr
    \times & \times & \triangle}
  \right),\nonumber \\
&&  {\bf E}^{}_5:\left(\matrix{
    \times & \triangle & \triangle \cr
    \triangle & 0 & \times \cr
    \triangle & \times & \times}
  \right), ~
  {\bf E}^{}_6:\left(\matrix{
    \times & \triangle & \times \cr
    \triangle & 0 & \triangle \cr
    \times & \triangle & \times}
  \right),~
  {\bf E}^{}_7:\left(\matrix{
    \times & \triangle & \times \cr
    \triangle & 0 & \times \cr
    \times & \times & \triangle}
  \right),~
  {\bf E}^{}_8:\left(\matrix{
    \times & \times & \triangle \cr
    \times & 0 & \triangle \cr
    \triangle & \triangle & \times}
  \right), \nonumber \\
&&  {\bf E}^{}_9:\left(\matrix{
    \times & \times & \triangle \cr
    \times & 0 & \times \cr
    \triangle & \times & \triangle}
  \right),~
  {\bf E}^{}_{10}:\left(\matrix{
    \times & \times & \times \cr
    \times & 0 & \triangle \cr
    \times & \triangle & \triangle}
  \right);
%     (5)
\end{eqnarray}
\vspace{0.2cm}
\begin{eqnarray}
&& {\bf F}^{}_1:\left(\matrix{
    \triangle & \triangle & \times \cr
    \triangle & \times & \times \cr
    \times & \times & 0}
    \right),~
{\bf F}^{}_2:\left(\matrix{
    \triangle & \times & \triangle \cr
    \times & \times & \times \cr
    \triangle & \times & 0}
  \right),~
{\bf F}^{}_3:\left(\matrix{
    \triangle & \times & \times \cr
    \times & \triangle & \times \cr
    \times & \times & 0}
  \right),~
{\bf F}^{}_4:\left(\matrix{
    \triangle & \times & \times \cr
    \times & \times & \triangle \cr
    \times & \triangle & 0}
  \right), \nonumber \\
&& {\bf F}^{}_5:\left(\matrix{
    \times & \triangle & \triangle \cr
    \triangle & \times & \times \cr
    \triangle & \times & 0}
  \right),~
{\bf F}^{}_6:\left(\matrix{
    \times & \triangle & \times \cr
    \triangle & \triangle & \times \cr
    \times & \times & 0}
  \right),~
{\bf F}^{}_7:\left(\matrix{
    \times & \triangle & \times \cr
    \triangle & \times & \triangle \cr
    \times & \triangle & 0}
  \right),~
{\bf F}^{}_8:\left(\matrix{
    \times & \times & \triangle \cr
    \times & \triangle & \times \cr
    \triangle & \times & 0}
  \right), \nonumber \\
&&  {\bf F}^{}_9:\left(\matrix{
    \times & \times & \triangle \cr
    \times & \times & \triangle \cr
    \triangle & \triangle & 0}
  \right),~
{\bf F}^{}_{10}:\left(\matrix{
    \times & \times & \times \cr
    \times & \triangle & \triangle \cr
    \times & \triangle & 0}
  \right) \; ,
%     (6)
\end{eqnarray}
where the triangles ``$\triangle$" denote equal and nonzero
elements, while the crosses ``$\times$" stand for arbitrary and
nonzero ones.

If one more element of $M^{}_\nu$ is assumed to be zero (or two more
elements are equal), the number of free parameters in $M^{}_\nu$
will be further reduced and the viable textures should be much less
(see, for example, Ref. \cite{Grimus}). However, we have numerically
checked that all those textures have already been excluded by
current neutrino oscillation data at the $3\sigma$ level.

\subsection{Important Relations}

In the basis where the charged-lepton mass matrix $M^{}_l$ is
diagonal, the Majorana neutrino mass matrix $M^{}_\nu$ can be
reconstructed from the leptonic mixing matrix $V$ and three neutrino
masses:
\begin{equation}
M^{}_\nu = V \left(\matrix{m^{}_1 & 0 & 0 \cr 0 & m^{}_2 & 0 \cr 0 &
0 & m^{}_3}\right) V^T \; .
%     (7)
\end{equation}
The leptonic mixing matrix can be parametrized as $V = U \cdot P$,
where the unitary matrix $U$ contains three mixing angles
$(\theta^{}_{12}, \theta^{}_{23}, \theta^{}_{13})$ and one
Dirac-type CP-violating phase $\delta$, namely,
\begin{equation}
U=\left( \matrix{c^{}_{12} c^{}_{13} & s^{}_{12} c^{}_{13}  &
s^{}_{13} \cr -c^{}_{12} s^{}_{23} s^{}_{13} - s^{}_{12} c^{}_{23}
e^{-i \delta } & -s^{}_{12} s^{}_{23} s^{}_{13} + c^{}_{12}
c^{}_{23} e^{-i \delta } &  s^{}_{23} c^{}_{13} \cr -c^{}_{12}
c^{}_{23} s^{}_{13} + s^{}_{12} s^{}_{23} e^{-i \delta }  &
-s^{}_{12} c^{}_{23} s^{}_{13}- c^{}_{12} s^{}_{23} e^{-i \delta} &
c^{}_{23} c^{}_{13}} \right) \; ;
%     (8)
\end{equation}
and $P = {\rm Diag}\{e^{i\rho}, e^{i\sigma}, 1\}$ is a diagonal
matrix with $\rho$ and $\sigma$ being two Majorana-type CP-violating
phases. Here we have defined $s^{}_{ij} \equiv \sin \theta^{}_{ij}$
and $c^{}_{ij} \equiv \cos \theta^{}_{ij}$ for $ij = 12, 23, 13$.
For later convenience, we rewrite
\begin{equation}
M^{}_\nu = U \left(\matrix{\lambda^{}_1 & 0 & 0 \cr 0 & \lambda^{}_2
& 0 \cr 0 & 0 & \lambda^{}_3}\right) U^T \; ,
%     (9)
\end{equation}
where $\lambda^{}_1 \equiv m^{}_1 e^{2i\rho}$, $\lambda^{}_2 \equiv
m^{}_2 e^{2i\sigma}$ and $\lambda^{}_3 \equiv m^{}_3$.

If one matrix element is zero [e.g., $(M^{}_\nu)^{}_{ab} = 0$] and
two other elements are equal [e.g., $(M^{}_\nu)^{}_{\alpha \beta} =
(M^{}_\nu)^{}_{cd}$], where three different independent elements of
$M^{}_\nu$ are considered, we obtain
\begin{eqnarray}
~~~~~~~~~~~~~~~~ \sum_{i=1}^3 U^{}_{ai} U^{}_{bi} \lambda^{}_i = 0
~~ {\rm and} ~~ \sum_{i=1}^3 (U^{}_{\alpha i} U^{}_{\beta i} -
U^{}_{c i} U^{}_{d i}) \lambda^{}_i = 0 \; ,
%     (10)
\end{eqnarray}
which lead to
\begin{eqnarray}
  \frac{\lambda^{}_1}{\lambda^{}_3} &=& \frac{U^{}_{a3} U^{}_{b3} U^{}_{\alpha 2} U^{}_{\beta 2}
  - U^{}_{a2} U^{}_{b2} U^{}_{\alpha 3} U^{}_{\beta 3} + U^{}_{a2} U^{}_{b2} U^{}_{c3} U^{}_{d3}
  - U^{}_{a3} U^{}_{b3} U^{}_{c2} U^{}_{d2}}{U^{}_{a2} U^{}_{b2} U^{}_{\alpha 1} U^{}_{\beta 1}
  - U^{}_{a1} U^{}_{b1} U^{}_{\alpha 2} U^{}_{\beta 2} + U^{}_{a1} U^{}_{b1} U^{}_{c2} U^{}_{d2}
  - U^{}_{a2} U^{}_{b2} U^{}_{c1} U^{}_{d1}} \; , \nonumber \\
  \frac{\lambda_2}{\lambda_3}&=&\frac{U^{}_{a1} U^{}_{b1} U^{}_{\alpha 3} U^{}_{\beta 3}
  - U^{}_{a3} U^{}_{b3} U^{}_{\alpha 1} U^{}_{\beta 1} + U^{}_{a3} U^{}_{b3} U^{}_{c1} U^{}_{d1}
  - U^{}_{a1} U^{}_{b1} U^{}_{c3} U^{}_{d3}}{U^{}_{a2} U^{}_{b2} U^{}_{\alpha 1} U^{}_{\beta 1}
  - U^{}_{a1} U^{}_{b1} U^{}_{\alpha 2} U^{}_{\beta 2} + U^{}_{a1} U^{}_{b1} U^{}_{c2} U^{}_{d2}
  - U^{}_{a2} U^{}_{b2} U^{}_{c1} U^{}_{d1}} \; .
  \label{eps.lambda}
%     (11)
\end{eqnarray}
With the help of Eq.~(\ref{eps.lambda}), one can figure out two
neutrino mass ratios $\xi \equiv m^{}_1/m^{}_3 =
|\lambda^{}_1/\lambda^{}_3|$ and $\zeta \equiv m^{}_2/m^{}_3 =
|\lambda^{}_2/\lambda^{}_3|$, as well as two Majorana CP-violating
phases $\rho = \arg[\lambda^{}_1/\lambda^{}_3]/2$ and $\sigma =
\arg[\lambda^{}_2/\lambda^{}_3]/2$. As we shall show later, these
important relations are quite useful in the determination of both
neutrino mass spectrum and leptonic CP-violating phases from current
experimental observations.

If both $(\theta^{}_{12}, \theta^{}_{23}, \theta^{}_{13})$ and
$\delta$ are precisely measured in neutrino oscillation experiments,
the unitary matrix $U$ is fixed, thus both $(\xi, \zeta)$ and
$(\rho, \sigma)$ can be determined from Eq.~(11). The neutrino mass
ratios $\xi$ and $\zeta$ are related to the ratio of two independent
neutrino mass-squared differences as
\begin{equation}
R^{}_\nu \equiv \frac{\delta m^2}{|\Delta m^2|} = \frac{2(\zeta^2 -
\xi^2)}{|2-(\zeta^2+\xi^2)|} \; ,
%      (12)
\end{equation}
and to three neutrino masses as
\begin{equation}
m^{}_3 = \sqrt{\frac{\delta m^2}{\zeta^2 - \xi^2}}\; , ~~~~ m^{}_2 =
m^{}_3 \zeta \; ,~~~~ m^{}_1 = m^{}_3 \xi \; .
%     (13)
\end{equation}
At present, the CP-violating phase $\delta$ is essentially
unconstrained in neutrino oscillation experiments. If one of the
hybrid textures is assumed, and three neutrino mixing angles and two
neutrino mass-squared differences are given, then $\delta$ can be
predicted from Eq.~(12). For the normal neutrino mass hierarchy, the
latest global-fit analysis yields at the $3 \sigma$ level
\cite{Fogli}
\begin{eqnarray}
0.259 \leq &\sin^2 \theta^{}_{12}& \leq 0.359 \; ,
\nonumber \\
0.331 \leq &\sin^2 \theta^{}_{23}& \leq 0.637 \; ,
\nonumber \\
0.017 \leq &\sin^2 \theta^{}_{13}& \leq 0.031 \; ,
%     (14)
\end{eqnarray}
and
\begin{eqnarray}
6.99\times 10^{-5}~{\rm eV}^2 \leq &\delta m^2& \leq 8.18 \times
10^{-5}~{\rm eV}^2 \; ,
\nonumber \\
2.19\times 10^{-3}~{\rm eV}^2 \leq & \hspace{-0.25cm}  \Delta m^2
\hspace{-0.25cm} & \leq 2.62 \times 10^{-3}~{\rm eV}^2 \; .
%     (15)
\end{eqnarray}
For the inverted neutrino mass hierarchy, the $3\sigma$ ranges of
neutrino mixing angles and mass-squared differences are slightly
different, so we shall use the same values as in the case of normal
neutrino mass hierarchy. In Table 1, the best-fit values together
with the $1\sigma$, $2\sigma$, and $3\sigma$ ranges are summarized
\footnote{The global-fit analysis of current neutrino oscillation
experiments has also been performed by two other groups
\cite{Schwetz,Valle}. Although their best-fit results of three
flavor mixing angles are slightly different from those obtained in
Ref. \cite{Fogli}, such differences become insignificant at the
$3\sigma$ level.}.

%%%%%%%%%%%%%%%%%%%%%%%%%%%%%%%%%%% Table 1 %%%%%%%%%%%%%%%%%%%%%%%%
\begin{table}[t]
\caption{The latest global-fit results of three neutrino mixing
angles $(\theta^{}_{12}, \theta^{}_{23}, \theta^{}_{13})$ and two
neutrino mass-squared differences $\delta m^2 \equiv m^2_2 - m^2_1$
and $\Delta m^2 \equiv m^2_3 - (m^2_1 + m^2_2)/2$ in the case of
normal neutrino mass hierarchy \cite{Fogli}.}
\begin{center}
\begin{tabular}{cccccc}
  \hline
  \hline
  Parameter & $\delta m^2~(10^{-5}~{\rm eV}^2)$ & $\Delta m^2~(10^{-3}~{\rm eV}^2)$
  & $\theta^{}_{12}$ & $\theta^{}_{23}$ & $\theta^{}_{13}$ \\
  \hline
  Best fit & $7.54$ & $2.43$ & $33.6^\circ$ & $38.4^\circ$ & $8.9^\circ$ \\
  $1\sigma$ range & $[7.32, 7.80]$ & $[2.33, 2.49]$ & $[32.6^\circ, 34.8^\circ]$
  & $[37.2^\circ, 40.0^\circ]$ & $[8.5^\circ, 9.4^\circ]$ \\
  $2\sigma$ range & $[7.15, 8.00]$ & $[2.27, 2.55]$ & $[31.6^\circ, 35.8^\circ]$
  & $[36.2^\circ, 42.0^\circ]$ & $[8.0^\circ, 9.8^\circ]$ \\
  $3\sigma$ range & $[6.99, 8.18]$ & $[2.19, 2.62]$ & $[30.6^\circ, 36.8^\circ]$
  & $[35.1^\circ, 53.0^\circ]$ & $[7.5^\circ, 10.2^\circ]$ \\
  \hline
\end{tabular}
\end{center}
\end{table}
%%%%%%%%%%%%%%%%%%%%%%%%%%%%%%%%%%%%%%%%%%%%%%%%%%%%%%%%%%%%%%%%%%%%%

\subsection{Quantum Corrections}

The stability of texture zeros in $M^{}_\nu$ against radiative
corrections has been extensively studied in the literature
\cite{FXZ}. By using the one-loop renormalization group equation
(RGE), one can demonstrate that the texture zeros in $M^{}_\nu$ at a
high-energy scale remain there at low-energy scales. In this
subsection, we examine the stability of texture zeros and equality
between two matrix elements against quantum corrections.

To be explicit, we consider the unique dimension-5 Weinberg operator
of massive Majorana neutrinos in an effective field theory after the
heavy degrees of freedom are integrated out \cite{Weinberg}:
\begin{equation}
\frac{{\cal L}^{}_{\rm d =5}}{\Lambda} = \frac{1}{2}
\kappa^{}_{\alpha \beta} \overline{\ell^{}_{\alpha \rm L}} \tilde{H}
\tilde{H}^T \ell^c_{\beta \rm L} + {\rm h.c.} \; ,
%     (16)
\end{equation}
where $\Lambda$ is the cutoff scale, $\ell^{}_{\rm L}$ denotes the
left-handed lepton doublet, $\tilde{H} \equiv i\sigma^{}_2 H^*$ with
$H$ being the standard-model Higgs doublet, and $\kappa$ stands for
the effective neutrino coupling matrix. After spontaneous gauge
symmetry breaking, $\tilde{H}$ gains its vacuum expectation value
$\langle \tilde{H} \rangle = v/\sqrt{2}$ with $v \approx 246$ GeV.
We are then left with the effective Majorana mass matrix $M^{}_\nu =
\kappa v^2/2$ for three light neutrinos from Eq.~(16). If the
dimension-5 Weinberg operator is obtained in the framework of the
minimal supersymmetric standard model, one will be left with
$M^{}_\nu = \kappa (v \sin\beta)^2/2$, where $\tan\beta$ denotes the
ratio of the vacuum expectation values of two Higgs doublets.
Eq.~(16) or its supersymmetric counterpart can provide a simple but
generic way of generating tiny neutrino masses. There are a number
of interesting possibilities of building renormalizable gauge models
to realize the effective Weinberg mass operator, such as the
well-known seesaw mechanisms at a superhigh energy scale $\Lambda$
\cite{SS1,SS2,SS3}.

The running of $M^{}_\nu$ from $\Lambda$ to the electroweak scale
$\mu \simeq M^{}_Z$ (or vice versa) is described by the RGE's
\cite{RGE}. In the chosen flavor basis and at the one-loop level,
$M^{}_\nu (M^{}_Z)$ and $M^{}_\nu (\Lambda)$ are related to each
other via
\begin{eqnarray}
M^{}_\nu (M^{}_Z) = I^{}_0 \left(\matrix{ I^{}_e & 0 & 0 \cr 0 &
I^{}_\mu & 0 \cr 0 & 0 & I^{}_\tau \cr} \right) M^{}_\nu (\Lambda)
\left(\matrix{ I^{}_e & 0 & 0 \cr 0 & I^{}_\mu & 0 \cr 0 & 0 &
I^{}_\tau \cr} \right) \; ,
%     (17)
\end{eqnarray}
where the RGE evolution function $I^{}_0$ denotes the overall
contribution from gauge and quark Yukawa couplings, and
$I^{}_\alpha$ (for $\alpha = e, \mu, \tau$) stand for the
contributions from charged-lepton Yukawa couplings \cite{Mei}.
Because of $I^{}_e < I^{}_\mu < I^{}_\tau$ as a consequence of
$m^{}_e \ll m^{}_\mu \ll m^{}_\tau$, they can modify the texture of
$M^{}_\nu$. In comparison, $I^{}_0 \neq 1$ only affects the overall
mass scale of $M^{}_\nu$. Note, however, that the texture zeros of
$M^{}_\nu$ are stable against such quantum corrections induced by
the one-loop RGE's. Taking {\bf Pattern} ${\bf A}^{}_1$ of
$M^{}_\nu$ for example, we have
\begin{eqnarray}
M^{{\bf A}^{}_1}_\nu (\Lambda) = \left(\matrix{ 0 & a & b \cr a & d
& d \cr b & d & c \cr} \right)
%     (18)
\end{eqnarray}
at $\Lambda$, and thus
\begin{eqnarray}
M^{{\bf A}^{}_1}_\nu (M^{}_Z) = I^{}_0 \left(\matrix{ 0 & a I^{}_e
I^{}_\mu & b I^{}_e I^{}_\tau \cr a I^{}_e I^{}_\mu & d I^2_\mu & d
I^{}_\mu I^{}_\tau \cr b I^{}_e I^{}_\tau & d I^{}_\mu I^{}_\tau & c
I^2_\tau \cr} \right)
%     (19)
\end{eqnarray}
at $M^{}_Z$. Although the texture zero is stable, the equality
$(M^{{\bf A}^{}_1}_\nu)^{}_{\mu \mu} = (M^{{\bf
A}^{}_1}_\nu)^{}_{\mu \tau}$ at $\Lambda$ is spoiled at the weak
scale $M^{}_Z$. Nevertheless, it can be shown that $I^{}_\alpha
\approx 1$ (for $\alpha = e, \mu, \tau$) hold as an excellent
approximation in the standard model. This interesting feature
implies that the important relations obtained in Eq.~(11) hold
approximately both at $\Lambda$ and $M^{}_Z$. In other words, if a
seesaw or flavor symmetry model predicts a hybrid texture of
$M^{}_\nu$ at $\Lambda$, one may simply study its phenomenological
consequences at $M^{}_Z$ by taking account of the same texture zero
and equality. However, the absolute values of neutrino masses are
indeed changed when running from a high-energy scale to low-energy
scales.

\section{Analytical Approximations}

First of all, we point out that there exists a permutation symmetry,
which relates one texture to another in Eqs.~(1)--(6). More
explicitly, the permutation between 2- and 3-rows of $M^{}_\nu$, and
that between 2- and 3-columns at the same time, change the position
of one zero and two equal elements, giving rise to another hybrid
texture $\tilde{M}^{}_\nu$. If $M^{}_\nu$ can be diagonalized by a
unitary matrix $U$ with mixing parameters $(\theta^{}_{12},
\theta^{}_{23}, \theta^{}_{13}, \delta)$, while $\tilde{M}^{}_\nu$
by a unitary matrix $\tilde{U}$ with mixing parameters
$(\tilde{\theta}^{}_{12}, \tilde{\theta}^{}_{23},
\tilde{\theta}^{}_{13}, \tilde{\delta})$, it is straightforward to
show that these two sets of mixing parameters are related as follows
\cite{FXZ}:
\begin{equation}
\tilde{\theta}^{}_{12} = \theta^{}_{12} \; , ~~~
\tilde{\theta}^{}_{13} = \theta^{}_{13} \; , ~~~
\tilde{\theta}^{}_{23} = \frac{\pi}{2} - \theta^{}_{23} \; , ~~~
\tilde{\delta} = \pi - \delta \; .
%     (20)
\end{equation}
Moreover, $M^{}_\nu$ and $\tilde{M}^{}_\nu$ have the same
eigenvalues $\lambda^{}_i$ (for $i = 1, 2, 3$). Among 39 viable
patterns, one can immediately verify that such a permutation
symmetry exists between
\begin{eqnarray}
&& {\bf A}^{}_1 \leftrightarrow {\bf A}^{}_3 \; , ~~~ {\bf B}^{}_1
\leftrightarrow {\bf C}^{}_1 \; , ~~~ {\bf B}^{}_2 \leftrightarrow
{\bf C}^{}_3 \; , ~~~ {\bf B}^{}_3 \leftrightarrow {\bf C}^{}_2 \; ,
~~~ {\bf B}^{}_4 \leftrightarrow {\bf C}^{}_5 \; ,\nonumber \\
&& {\bf B}^{}_5 \leftrightarrow {\bf C}^{}_4 \; , ~~~ {\bf D}^{}_1
\leftrightarrow {\bf D}^{}_2 \; , ~~~ {\bf D}^{}_3 \leftrightarrow
{\bf D}^{}_6 \; , ~~~ {\bf D}^{}_4 \leftrightarrow {\bf D}^{}_5 \; ,
~~~ {\bf E}^{}_1 \leftrightarrow {\bf F}^{}_2 \; , \nonumber \\
&& {\bf E}^{}_2 \leftrightarrow {\bf F}^{}_1 \; , ~~~~ {\bf E}^{}_3
\leftrightarrow {\bf F}^{}_4 \; , ~~~~ {\bf E}^{}_4 \leftrightarrow
{\bf F}^{}_3 \; , ~~~ {\bf E}^{}_5 \leftrightarrow {\bf F}^{}_5 \; ,
~~~ {\bf E}^{}_6 \leftrightarrow {\bf F}^{}_9 \; , \nonumber \\
&& {\bf E}^{}_7 \leftrightarrow {\bf F}^{}_8 \; , ~~~~ {\bf E}^{}_8
\leftrightarrow {\bf F}^{}_7 \; , ~~~~ {\bf E}^{}_9 \leftrightarrow
{\bf F}^{}_6 \; , ~~~ {\bf E}^{}_{10} \leftrightarrow {\bf
F}^{}_{10} \; ,
%     (21)
\end{eqnarray}
so the analytical results in Eq.~(11) for one hybrid texture can be
obtained from those for the corresponding paired one. Hence we are
left with only twenty independent patterns. It is worthwhile to
mention that ${\bf Pattern}~{\bf A}^{}_2$ is invariant under the
permutations of 2- and 3-rows and columns.

In the following, we focus on the approximate analytical results for
the six patterns ${\bf A}^{}_1$, ${\bf B}^{}_1$, ${\bf B}^{}_5$,
${\bf D}^{}_1$, ${\bf E}^{}_1$, and ${\bf E}^{}_8$ and explore their
implications for neutrino mass spectrum and the leptonic
CP-violating phases. The detailed numerical studies will be given in
section 4. The analytical approximations for the other patterns can
be discussed in a similar way, but they are more or less dependent
on whether the mixing angle $\theta^{}_{23}$ is close to $\pi/4$ and
whether the Dirac CP-violating phase $\delta$ is nearly $\pi/2$.
Another important motivation to choose these six patterns for
illustration is that they provide very concrete predictions either
for the mixing angles, or for the neutrino mass hierarchy, or for
the Dirac CP-violating phase, or for the neutrinoless double-beta
decays, which make them phenomenologically more interesting and
experimentally more testable than the other patterns.
\begin{itemize}
\item {${\bf Pattern}~{\bf A}^{}_1$} with $(M^{}_\nu)^{}_{ee} = 0$
and $(M^{}_\nu)^{}_{\mu\mu} = (M^{}_\nu)^{}_{\mu\tau}$. With the
help of Eq.~(\ref{eps.lambda}), in the leading order of $\sin
\theta^{}_{13}$, we have
\begin{eqnarray}
\frac{\lambda^{}_1}{\lambda^{}_3} &\approx& - \frac{\sin^2
\theta^{}_{12} \left(1 - \tan\theta^{}_{23}\right)} {\cos
2\theta^{}_{12} (1 + \cot \theta^{}_{23})}~e^{2 i \delta} \; ,
\nonumber \\
\frac{\lambda^{}_2}{\lambda^{}_3} &\approx& + \frac{\cos^2
\theta^{}_{12} \left(1 - \tan\theta^{}_{23}\right)} {\cos
2\theta^{}_{12} (1 + \cot \theta^{}_{23})}~e^{2 i \delta} \; ,
%     (22)
\end{eqnarray}
which lead us to the neutrino mass ratios
\begin{eqnarray}
\xi &\approx& \frac{\sin^2 \theta^{}_{12} |1-\tan\theta _{23}|}
{\cos 2\theta^{}_{12} (1 + \cot \theta^{}_{23})} \; , \nonumber\\
\zeta &\approx& \frac{\cos^2 \theta^{}_{12} |1-\tan\theta _{23}|}
{\cos 2\theta^{}_{12} (1 + \cot \theta^{}_{23})} \; ,
%     (23)
\end{eqnarray}
and the relations between Majorana and Dirac CP-violating phases:
$\rho \approx \delta - \pi/2$ and $\sigma \approx \delta$ (for
$\theta^{}_{23} < 45^\circ$) or $\rho \approx \delta$ and $\sigma
\approx \delta - \pi/2$ (for $\theta^{}_{23} > 45^\circ$).

Taking the $3\sigma$ ranges of neutrino mixing angles, we obtain
$0.59 \leq \tan \theta^{}_{12} \leq 0.75$ and $0.70 \leq \tan
\theta^{}_{23} \leq 1.3$, which yield $\xi < \zeta <1$. Therefore,
only the normal neutrino mass hierarchy $m^{}_1 < m^{}_2 < m^{}_3$
or equivalently $\Delta m^2 > 0$ is allowed. Furthermore, we get
\begin{eqnarray}
R^{}_\nu \approx \zeta^2 - \xi^2 \approx \sec 2\theta^{}_{12}
\left(\frac{1 - \tan \theta^{}_{23}} {1 + \cot
\theta^{}_{23}}\right)^2,
%     (24)
\end{eqnarray}
which is actually independent of $\delta$. In order to determine or
constrain $\delta$, we have to work in the next-to-leading order
approximation. More explicitly, we obtain
\begin{eqnarray}
\xi &\approx& \frac{\sin^2 \theta^{}_{12} |1 - \tan
\theta^{}_{23}|}{\cos 2\theta^{}_{12} (1 + \cot \theta^{}_{23})}
\left(1 + \frac{1-\cot 2\theta^{}_{23}}{1 + \cot \theta^{}_{23}}
\tan 2\theta^{}_{12} \sin \theta^{}_{13} \cos \delta\right) \; ,
\nonumber \\
\zeta &\approx& \frac{\cos^2 \theta^{}_{12} |1 - \tan
\theta^{}_{23}|}{\cos 2\theta^{}_{12} (1 + \cot \theta^{}_{23})}
\left(1 + \frac{1 - \cot 2\theta^{}_{23}} {1 + \cot \theta^{}_{23}}
\tan 2\theta^{}_{12} \sin \theta^{}_{13} \cos \delta \right) \; ,
%     (25)
\end{eqnarray}
and thus
\begin{eqnarray}
R^{}_\nu \approx \sec 2\theta^{}_{12} \left(\frac{1 - \tan
\theta^{}_{23}} {1+\cot\theta _{23}}\right)^2 \left[1 + 2 \tan
2\theta^{}_{12} \sin \theta^{}_{13} \cos \delta \left(\frac{1 - \cot
2\theta^{}_{23}} {1 + \cot \theta^{}_{23}}\right)\right].
%     (26)
\end{eqnarray}
Now it is straightforward to solve Eq.~(26) for the CP-violating
phase, namely,
\begin{eqnarray}
\delta &\approx& \cos^{-1} \left\{\frac{\cot2
\theta_{12}(1+\cot\theta_{23})} {2\sin\theta^{}_{13} \left(1 - \cot
2\theta^{}_{23}\right)} \left[\frac{R^{}_\nu (1 + \cot
\theta^{}_{23})^2}{\sec 2\theta^{}_{12} (1 - \tan \theta^{}_{23})^2}
- 1\right]\right\}.
%     (27)
\end{eqnarray}
For the best-fit values of $(\theta^{}_{12}, \theta^{}_{23},
\theta^{}_{13})$ and $R^{}_\nu$ from Table 1, there is no solution
to Eq.~(27). Taking $\theta^{}_{12} = 35^\circ$ and setting the
other parameters to their best-fit values (i.e., $\theta^{}_{23} =
38.4^\circ, \theta^{}_{13} = 8.9^\circ, \delta m^2 = 7.54 \times
10^{-5}~\eV^2$ and $\Delta m^2 = 2.43 \times 10^{-3}~\eV^2$), one
can figure out the Dirac and Majorana CP-violating phases
\begin{eqnarray}
\delta \approx 23^\circ \; , ~~~ \rho \approx -67^\circ \; , ~~~
\sigma \approx 23^\circ \; ,
%     (28)
\end{eqnarray}
as well as the neutrino mass spectrum
\begin{eqnarray}
m^{}_3 &\approx& \sqrt{\Delta m^2} \approx 4.9 \times 10^{-2}~\eV \; ,
\nonumber \\
m^{}_2 &\approx& m^{}_3 \zeta \approx 8.9 \times 10^{-3}~\eV \; ,
\nonumber\\
m^{}_1 &\approx& m^{}_3 \xi \approx 4.3 \times 10^{-3}~\eV \; .
%     (29)
\end{eqnarray}
Since $(M^{}_\nu)^{}_{ee} = 0$ holds for {\bf Pattern} ${\bf
A}^{}_1$, the effective neutrino mass $\langle m \rangle^{}_{\rm
ee}$ in the neutrinoless double-beta ($0\nu2\beta$) decays is
vanishing. The future observation of $0\nu2\beta$ decays will
definitely rule out this pattern. See, e.g., Ref.~\cite{Rodejohann},
for recent reviews on the theoretical and experimental status of the
$0\nu2\beta$ decays.

\item {\bf Pattern} ${\bf B}^{}_1$ with $(M^{}_\nu)^{}_{e\mu}=0$ and
$(M^{}_\nu)^{}_{ee} = (M^{}_\nu)^{}_{e\tau}$. In the leading order
of $\sin \theta^{}_{13}$, one can obtain from Eq.~(11) that
\begin{eqnarray}
\frac{\lambda^{}_1}{\lambda^{}_3} &\approx& \frac{\sin
\theta^{}_{13}}{\cos \theta^{}_{23}} \left(1 + \tan \theta^{}_{12}
\sin \theta^{}_{23} e^{i \delta}\right) \; , \nonumber \\
\frac{\lambda^{}_2}{\lambda^{}_3} &\approx& \frac{\sin
\theta^{}_{13}}{\cos \theta^{}_{23}} \left(1 - \cot \theta^{}_{12}
\sin \theta^{}_{23} e^{i \delta}\right) \; ,
%     (30)
\end{eqnarray}
leading to the neutrino mass ratios
\begin{eqnarray}
\xi &\approx& \frac{\sin \theta^{}_{13}}{\cos \theta^{}_{23}}
\left(1 + \tan^2 \theta^{}_{12} \sin^2 \theta^{}_{23} + 2 \tan
\theta^{}_{12} \sin \theta^{}_{23} \cos \delta\right)^{1/2} \; ,
\nonumber \\
\zeta &\approx& \frac{\sin \theta^{}_{13}}{\cos \theta^{}_{23}}
\left(1 + \cot^2 \theta^{}_{12} \sin^2 \theta^{}_{23} - 2 \cot
\theta^{}_{12} \sin \theta^{}_{23} \cos \delta\right)^{1/2} \; ,
%     (31)
\end{eqnarray}
and the Majorana CP-violating phases
\begin{eqnarray}
\rho &\approx& \frac{1}{2} \arg \left(1 + \tan \theta^{}_{12}
\sin \theta^{}_{23} e^{i \delta}\right) \; , \nonumber\\
\sigma &\approx& \frac{1}{2} \arg \left(1 - \cot \theta^{}_{12} \sin
\theta^{}_{23} e^{i \delta}\right) \; .
%     (32)
\end{eqnarray}
Taking the values of three neutrino mixing angles $(\theta^{}_{12},
\theta^{}_{23}, \theta^{}_{13})$ within their $3\sigma$ ranges, one
can verify $\xi < \zeta<1$, implying that only the normal neutrino
mass hierarchy ($m^{}_1 < m^{}_2 < m^{}_3$) is allowed. Furthermore,
we obtain
\begin{eqnarray}
R^{}_\nu \approx \zeta^2 - \xi^2 \approx 4\sin^2 \theta^{}_{13} \csc
2\theta^{}_{12} \tan^2 \theta^{}_{23}(\cot2\theta^{}_{12} - \csc
\theta^{}_{23} \cos\delta),
%     (33)
\end{eqnarray}
from which one can determine the CP-violating phase
\begin{eqnarray}
\delta\approx\cos^{-1}\left[\sin\theta _{23}\left(\cot2\theta _{12}
- \frac{R_\nu \sin 2\theta _{12}}{4\sin^2 \theta _{13} \tan^2
\theta_{23}}\right)\right].
%     (34)
\end{eqnarray}
Since neutrino oscillation experiments indicate $m^{}_1 < m^{}_2$
(i.e., $R^{}_\nu > 0$), the condition $\cos \delta <
\cot2\theta^{}_{12} \sin\theta^{}_{23}$ should be satisfied. With
the best-fit values of three neutrino mixing angles and two neutrino
mass-squared differences given in Table 1, we arrive at
\begin{eqnarray}
\delta \approx 92^\circ \; , ~~~~ \rho \approx 11^\circ \; , ~~~~
\sigma \approx -21^\circ,
%     (35)
\end{eqnarray}
while three neutrino masses are
\begin{eqnarray}
m^{}_3 &\approx& \sqrt{|\Delta m^2|} \approx 4.9\times10^{-2}~\eV \; ,
\nonumber\\
m^{}_2 &\approx& m^{}_3 \zeta \approx 1.4\times10^{-2}~\eV \; ,
\nonumber\\
m^{}_1 &\approx& m^{}_3 \xi \approx 1.0\times10^{-2}~\eV \; .
%     (36)
\end{eqnarray}
In addition, the effective neutrino mass in the $0\nu2\beta$ decays
can be estimated as $\langle m \rangle_{\rm ee} \approx m_3
\sin\theta _{13} \sec\theta_{23} \approx 9.8 \times 10^{-3}~\eV$,
which is quite challenging for the experimental searches even at the
next-generation facilities \cite{Rodejohann}.

\item {\bf Pattern} ${\bf B}^{}_5$ with $(M^{}_\nu)^{}_{e\mu}=0$ and
$(M^{}_\nu)^{}_{\mu\tau} = (M^{}_\nu)^{}_{\tau\tau}$. With the help
of Eq. (\ref{eps.lambda}), in the leading order of
$\sin\theta^{}_{13}$, we get
\begin{eqnarray}
\frac{\lambda^{}_1}{\lambda^{}_3} \approx
\frac{\lambda^{}_2}{\lambda^{}_3} \approx \frac{e^{2 i \delta }
\left(1 - \cot\theta^{}_{23}\right)}{1 + \tan\theta^{}_{23}} \; ,
%     (37)
\end{eqnarray}
from which follows
\begin{eqnarray}
\xi \approx \zeta \approx \frac{|1-\cot\theta^{}_{23}|} {1 +
\tan\theta^{}_{23}} \; ,
%     (38)
\end{eqnarray}
and $\rho \approx \sigma \approx \delta -\pi/2$ (for $\theta^{}_{23}
< 45^\circ$) or $\rho \approx \sigma \approx \delta$ (for
$\theta^{}_{23} > 45^\circ$). Since $R^{}_\nu \approx 0$ in the
leading-order approximation, we have to work at the next-to-leading
order
\begin{eqnarray}
\frac{\lambda^{}_1}{\lambda^{}_3} &\approx& \frac{e^{2 i \delta }
\left(1-\cot\theta^{}_{23}\right)}{1+\tan\theta^{}_{23}} \left[1 +
\frac{\sin\theta^{}_{13} \cos\delta \left[1 - i \sqrt{2} \tan \delta
\sin (2\theta^{}_{23} -\pi/4)\right]}
{\tan\theta^{}_{12} \cos^2\theta^{}_{23}(1 - \cot\theta^{}_{23})}\right]
\; , \nonumber \\
\frac{\lambda^{}_2}{\lambda^{}_3} &\approx& \frac{e^{2 i \delta }
\left(1 - \cot\theta^{}_{23}\right)}{1 + \tan\theta^{}_{23}} \left[1
- \frac{\sin\theta^{}_{13} \cos \delta \left[1 - i \sqrt{2}
\tan\delta \sin (2\theta^{}_{23} -\pi/4)\right]} {\cot
\theta^{}_{12} \cos^2\theta^{}_{23}(1 - \cot\theta^{}_{23})}\right]
\; .
%      (39)
\end{eqnarray}
Hence, to the first order of $\sin \theta^{}_{13}$, one obtains
\begin{eqnarray}
R^{}_\nu \approx \zeta^2 - \xi^2 \approx \frac{2(1 -
\tan\theta^{}_{23}) \sin\theta^{}_{13} \cos\delta} {(1 -
\cot2\theta^{}_{23}) \sin 2\theta^{}_{12}} \; ,
%     (40)
\end{eqnarray}
leading to
\begin{eqnarray}
\delta &\approx& \cos^{-1} \left[ \frac{(1 - \cot2\theta _{23})
\sin2\theta^{}_{12} R^{}_\nu} {2(1 - \tan\theta^{}_{23})
\sin\theta^{}_{13}}\right] \; ,
%     (41)
\end{eqnarray}
and the difference between two neutrino mass ratios
\begin{eqnarray}
\zeta - \xi \approx \frac{(1 - \tan^2 \theta^{}_{23})
\sin\theta^{}_{13} \cos\delta}{(1 - \cot2\theta^{}_{23})|1 -
\cot\theta^{}_{23}| \sin2\theta^{}_{12}} \; .
%     (42)
\end{eqnarray}
Taking the values of $\theta^{}_{23}$ within the $3\sigma$ range, we
have $\xi \approx \zeta < 1$, implying that only the normal neutrino
mass hierarchy is allowed. To ensure $\xi < \zeta$ or equivalently
$m^{}_1 < m^{}_2$, we have to require $\cos \delta > 0$ for
$\theta^{}_{23} < 45^\circ$, and $\cos\delta < 0$ for
$\theta^{}_{23} > 45^\circ$. Furthermore, using the best-fit values
of three neutrino mixing angles and two neutrino mass-squared
differences, we arrive at
\begin{eqnarray}
\delta \approx 70^\circ \; , ~~~~ \rho \approx 12^\circ \; , ~~~~
\sigma \approx -30^\circ \; ,
%     (43)
\end{eqnarray}
and the neutrino mass spectrum
\begin{eqnarray}
m^{}_3 &\approx& (1 + \tan \theta^{}_{23}) \sqrt{\frac{\Delta m^2
\sin 2\theta^{}_{23}}{1 - \cot 2\theta^{}_{23}}} \approx 5.0 \times 10^{-2}~\eV
\; ,\nonumber\\
m^{}_1 &\approx& m^{}_2 ~~ \approx ~~ m^{}_3 \cdot \frac{|1 -
\cot\theta^{}_{23}|}{1 + \tan\theta^{}_{23}} \approx 7.3 \times
10^{-3}~\eV \; .
%     (44)
\end{eqnarray}
It is straightforward to calculate the effective neutrino mass
$\langle m\rangle^{}_{\rm ee} \approx 7.3 \times 10^{-3}~\eV$ in the
$0\nu2\beta$ decays. As expected for the case of normal neutrino
mass hierarchy, $\langle m \rangle^{}_{\rm ee}$ is too small to be
measured in the near future.

\item {\bf Pattern} ${\bf D}^{}_1$ with $(M^{}_\nu)^{}_{\mu\tau} = 0$
and $(M^{}_\nu)^{}_{ee} = (M^{}_\nu)^{}_{e\mu}$. From Eq.~(11), in
the leading order of $\sin\theta_{13}$, we derive
\begin{eqnarray}
\frac{\lambda^{}_1}{\lambda^{}_3} &\approx& e^{2 i \delta}
\frac{\cos\theta^{}_{23} - \tan\theta^{}_{12} e^{i \delta}}
{\cos\theta^{}_{23} + 2 \cot 2\theta^{}_{12} e^{i \delta}}
\; , \nonumber\\
\frac{\lambda^{}_2}{\lambda^{}_3} &\approx& e^{2 i \delta}
\frac{\cos\theta^{}_{23} + \cot\theta^{}_{12} e^{i
\delta}}{\cos\theta^{}_{23} + 2 \cot 2\theta^{}_{12} e^{i \delta}}
\; .
%     (45)
\end{eqnarray}
From Eq.~(45), it is straightforward to extract the neutrino mass
ratios
\begin{eqnarray}
\xi &\approx& \left[\frac{\cos^2\theta^{}_{23} +
\tan^2\theta^{}_{12} - 2\cos\delta \cos\theta^{}_{23}
\tan\theta^{}_{12}}{\cos^2\theta^{}_{23} + 4 \cot^2 2\theta^{}_{12}
+ 4\cos\delta \cot 2\theta^{}_{12} \cos\theta^{}_{23}}\right]^{1/2} \; , \nonumber\\
\zeta &\approx& \left[\frac{\cos^2 \theta^{}_{23} + \cot^2
\theta^{}_{12} + 2\cos\delta \cos\theta^{}_{23} \cot\theta^{}_{12}}
{\cos^2\theta^{}_{23} + 4 \cot^2 2\theta^{}_{12} + 4\cos\delta \cot
2\theta^{}_{12} \cos\theta^{}_{23}}\right]^{1/2} \; ,
%     (46)
\end{eqnarray}
and the Majorana CP-violating phases
\begin{eqnarray}
\rho &\approx& \delta + \frac{1}{2} \arg
\left[\frac{\cos\theta^{}_{23} - \tan\theta^{}_{12} e^{i \delta}}
{\cos\theta^{}_{23} + 2 \cot 2\theta^{}_{12} e^{i \delta}}\right] \;
,
\nonumber\\
\sigma &\approx & \delta + \frac{1}{2}\arg
\left[\frac{\cos\theta^{}_{23} + \cot\theta^{}_{12} e^{i \delta}}
{\cos\theta^{}_{23} + 2 \cot 2\theta^{}_{12} e^{i \delta}}\right] \;
.
%     (47)
\end{eqnarray}
Moreover, it is easy to show that $\cos \delta > - \cot
2\theta^{}_{12} \sec \theta^{}_{23}$ must be satisfied in order to
guarantee $\zeta > \xi$ or equivalently $m^{}_2 > m^{}_1$. Inputting
the neutrino mixing angles in their $3\sigma$ ranges and $\delta \in
[0, 2\pi)$, we find $\zeta > \xi > 1$, so only the inverted neutrino
mass hierarchy is possible. Thus we get
\begin{eqnarray}
R^{}_\nu = \frac{2(\zeta^2-\xi^2)}{(\zeta^2+\xi^2)-2} \approx -
\frac{8\left(\cos 2\theta^{}_{12} + \sin 2\theta^{}_{12}
\cos\theta^{}_{23} \cos\delta\right)}{1 + 3\cos 4\theta^{}_{12} + 2
\sin 4\theta^{}_{12} \cos\theta^{}_{23} \cos\delta} \; ,
%     (48)
\end{eqnarray}
which leads us to
\begin{eqnarray}
\delta \approx \cos^{-1} \left[ -\frac{8 \cos 2\theta^{}_{12}
+\left(1 + 3 \cos 4\theta^{}_{12}\right)
R^{}_\nu}{4\cos\theta^{}_{23} \sin 2\theta^{}_{12} \left(2 + \cos
2\theta^{}_{12} R^{}_\nu\right)}\right] \; .
%     (49)
\end{eqnarray}
With the best-fit values of three neutrino mixing angles and two
neutrino mass-squared differences, we obtain three CP-violating
phases
\begin{eqnarray}
\delta \approx 122^\circ \; , ~~~~ \rho \approx 76^\circ \; , ~~~~
\sigma \approx -45^\circ \; ,
%     (50)
\end{eqnarray}
and three neutrino masses
\begin{eqnarray}
m^{}_3 &\approx& \sqrt{\frac{\delta m^2}{\zeta^2 - \xi^2}} \approx
3.91 \times 10^{-2}~\eV \; , \nonumber\\
m^{}_2 &\approx& m^{}_3 \zeta \approx 6.32 \times 10^{-2}~\eV \; , \nonumber\\
m^{}_1 &\approx& m^{}_3 \zeta \approx 6.26 \times 10^{-2}~\eV \; .
%     (51)
\end{eqnarray}
Note that the neutrino mass spectrum is nearly degenerate, so the
effective neutrino mass in the $0\nu2\beta$ decays $\langle
m\rangle^{}_{\rm ee} \approx m^{}_3/|1 + 2 e^{i \delta} \cot 2\theta
_{12} \sec\theta _{23}| \approx 3.87 \times 10^{-2}~\eV$ turns out
to be accessible in the next-generation experiments.

\item {\bf Pattern} ${\bf E}^{}_1$ with $(M^{}_\nu)^{}_{\mu\mu} = 0$
and $(M^{}_\nu)^{}_{ee} = (M^{}_\nu)^{}_{e\mu}$. With the help of
Eq. (\ref{eps.lambda}), in the leading order of $\sin\theta _{13}$,
we have
\begin{eqnarray}
\frac{\lambda^{}_1}{\lambda^{}_3} &\approx& -e^{2 i\delta} \tan^2
\theta^{}_{23} \frac{\cos \theta^{}_{23} - \tan \theta^{}_{12} e^{i
\delta}}{\cos \theta^{}_{23} + 2 \cot 2\theta^{}_{12} e^{i \delta}}
\; ,\nonumber\\
\frac{\lambda^{}_2}{\lambda^{}_3} &\approx& -e^{2 i\delta} \tan^2
\theta^{}_{23} \frac{\cos \theta^{}_{23} + \cot \theta^{}_{12} e^{i
\delta}}{\cos \theta^{}_{23} + 2\cot 2\theta^{}_{12} e^{i\delta}}
\;.
%     (52)
\end{eqnarray}
From Eq.~(52), it is straightforward to extract the neutrino mass
ratios
\begin{eqnarray}
\xi &\approx& \tan^2 \theta^{}_{23} \left[\frac{\cos^2\theta^{}_{23}
+ \tan^2\theta^{}_{12} - 2\cos\delta \cos\theta^{}_{23}
\tan\theta^{}_{12}}{\cos^2\theta^{}_{23} + 4 \cot^2 2\theta^{}_{12}
+ 4\cos\delta \cot 2\theta^{}_{12} \cos\theta^{}_{23}}\right]^{1/2}
\; , \nonumber\\
\zeta &\approx& \tan^2 \theta^{}_{23} \left[\frac{\cos^2
\theta^{}_{23} + \cot^2 \theta^{}_{12} + 2\cos\delta
\cos\theta^{}_{23} \cot\theta^{}_{12}} {\cos^2\theta^{}_{23} + 4
\cot^2 2\theta^{}_{12} + 4\cos\delta \cot 2\theta^{}_{12}
\cos\theta^{}_{23}}\right]^{1/2} \; ,
%     (53)
\end{eqnarray}
and the Majorana CP-violating phases
\begin{eqnarray}
\rho &\approx& \delta + \frac{1}{2} \arg
\left[\frac{-\cos\theta^{}_{23} + \tan\theta^{}_{12} e^{i \delta}}
{\cos\theta^{}_{23} + 2 \cot 2\theta^{}_{12} e^{i \delta}}\right] \;
, \nonumber\\
\sigma &\approx & \delta + \frac{1}{2}\arg
\left[\frac{-\cos\theta^{}_{23} - \cot\theta^{}_{12} e^{i \delta}}
{\cos\theta^{}_{23} + 2 \cot 2\theta^{}_{12} e^{i \delta}}\right]
\;.
%     (54)
\end{eqnarray}
Combining Eq. (53) with Eq. (12), we can determine the Dirac
CP-violating phase
\begin{eqnarray}
\delta \approx \cos^{-1} \left\{-\frac{\cot 2\theta^{}_{12}} {\cos
\theta^{}_{23}} \left[1 + \frac{\tan^2 2\theta^{}_{12} \left( \cot^4
\theta^{}_{23} - \csc^2 \theta^{}_{23}\right) R^{}_\nu}{4\sec
2\theta^{}_{12} + 2\left( 2 \cot^4 \theta^{}_{23} - 1\right)
R^{}_\nu}\right]\right\} \; .
%     (55)
\end{eqnarray}
Taking the best-fit values of three neutrino mixing angles and two
neutrino mass-squared differences, we obtain
\begin{eqnarray}
\delta \approx 122^\circ \; , ~~~~ \rho \approx -13^\circ \; , ~~~~
\sigma \approx 45^\circ \; ,
%     (56)
\end{eqnarray}
and the neutrino mass spectrum is as follows
\begin{eqnarray}
m^{}_3 &\approx& \sqrt{\frac{\delta m^2}{\zeta^2 - \xi^2}}
\approx 9.24 \times 10^{-2}~\eV \; , \nonumber\\
m^{}_1 &\approx& m^{}_2 ~~\approx~~9.41 \times 10^{-2}~\eV \; .
%     (57)
\end{eqnarray}
Finally, it is straightforward to figure out $\langle
m\rangle^{}_{\rm ee} \approx m^{}_3 \tan^2 \theta^{}_{23}/|1 + 2e^{i
\delta} \cot 2\theta^{}_{12} \sec \theta^{}_{23}| \approx
5.8\times10^{-2}~\eV$, which is quite encouraging for the upcoming
$0\nu2\beta$ experiments.

It is worthwhile to mention that the analytical formulas for {\bf
Pattern} ${\bf D}^{}_1$ are identical to those for {\bf Pattern}
${\bf E}^{}_1$ if $\theta^{}_{23} = \pi/4$ is assumed.  Therefore,
the precision measurement of $\theta^{}_{23}$ is crucial to
distinguish between these two patterns of $M^{}_\nu$.

\item {\bf Pattern} ${\bf E}^{}_8$ with $(M^{}_\nu)^{}_{\mu\mu} = 0$
and $(M^{}_\nu)^{}_{e\tau} = (M^{}_\nu)^{}_{\mu\tau}$. From Eq.
(11), in the leading order of $\sin\theta_{13}$, we obtain
\begin{eqnarray}
\frac{\lambda^{}_1}{\lambda^{}_3} &\approx& -e^{2i\delta}
(\tan^2 \theta^{}_{23} - \cot \theta^{}_{12} \sec \theta^{}_{23}
e^{-i\delta}) \; , \nonumber\\
\frac{\lambda^{}_2}{\lambda^{}_3} &\approx& -e^{2i\delta} (\tan^2
\theta^{}_{23} + \tan \theta^{}_{12} \sec \theta^{}_{23} e^{-i
\delta}) \; .
%     (58)
\end{eqnarray}
Then, from Eq. (58), it is easy to extract the neutrino mass ratios
\begin{eqnarray}
\xi &\approx& \left[\tan^4 \theta^{}_{23} + \cot^2 \theta^{}_{12}
\sec^2 \theta^{}_{23} - 2\cot \theta^{}_{12} \tan^2 \theta^{}_{23}
\sec \theta^{}_{23} \cos\delta \right]^{1/2} \; , \nonumber\\
\zeta &\approx& \left[\tan^4 \theta^{}_{23} + \tan^2 \theta^{}_{12}
\sec^2 \theta^{}_{23} + 2\tan \theta^{}_{12} \tan^2 \theta^{}_{23}
\sec \theta^{}_{23} \cos\delta\right]^{1/2} \; ,
%     (59)
\end{eqnarray}
and the Majorana CP-violating phases
\begin{eqnarray}
\rho & \approx & \delta + \frac{1}{2} \arg \left[-\tan^2
\theta^{}_{23} + \cot \theta^{}_{12} \sec\theta^{}_{23}
e^{-i\delta}\right] \; ,\nonumber\\
\sigma &\approx & \delta + \frac{1}{2} \arg \left[-\tan^2
\theta^{}_{23} - \tan \theta^{}_{12} \sec\theta^{}_{23}
e^{-i\delta}\right] \; .
%     (60)
\end{eqnarray}
One immediately observes from Eq. (59) that
\begin{eqnarray}
\zeta^2 - \xi^2 &\approx& \frac{4(\cos\delta \tan^2 \theta^{}_{23} -
\cot 2\theta^{}_{12} \sec\theta^{}_{23})}{\sin 2\theta^{}_{12} \cos
\theta^{}_{23}} \; .
%     (61)
\end{eqnarray}
In order to ensure $m^{}_1 < m^{}_2$ or equivalently $\zeta^2 -
\xi^2 > 0$, we have to require $\cos \delta > \cot 2\theta^{}_{12}
\sec \theta^{}_{23} \cot^2 \theta^{}_{23} > 0$, implying $\delta <
\pi/2$ or $\delta > 3\pi/2$. Taking the $3\sigma$ ranges of the
mixing parameters, we find that only the inverted neutrino mass
hierarchy is allowed. Moreover, we get
\begin{eqnarray}
R^{}_\nu \approx \frac{4 \csc 2\theta^{}_{12} (\cos\delta \tan^2
\theta^{}_{23} - \cot 2\theta^{}_{12} \sec \theta^{}_{23})} {\left(2
\cot^2 2\theta^{}_{12} + \tan^2 \theta^{}_{23} \right) \sec
\theta^{}_{23} - 2 \cos\delta \cot 2\theta^{}_{12} \tan^2
\theta^{}_{23}} \; ,
%     (62)
\end{eqnarray}
from which one can determine the Dirac CP-violating phase
\begin{eqnarray}
\delta &\approx& \cos^{-1} \left[\frac{\cot 2\theta^{}_{12}}
{\tan\theta^{}_{23} \sin\theta^{}_{23}} + \frac{\sin 2\theta^{}_{12}
\sec\theta^{}_{23} R^{}_\nu}{4 + 2 \cos 2\theta^{}_{12} R^{}_\nu}
\right] \; .
%     (63)
\end{eqnarray}
With the best-fit values of three neutrino mixing angles and two
neutrino mass-squared differences, we arrive at
\begin{eqnarray}
\delta \approx 30^\circ \; , ~~~~ \rho \approx 9^\circ \; , ~~~~
\sigma \approx -68^\circ \; .
%     (64)
\end{eqnarray}
In addition, three neutrino masses are found to be
\begin{eqnarray}
m^{}_3 &\approx& \left[\frac{\delta m^2 \sin 2\theta^{}_{12} \cos
\theta^{}_{23}} {4(\cos\delta \tan^2\theta^{}_{23} - \cot
2\theta^{}_{12} \sec \theta^{}_{23})}\right]^{1/2} \approx 3.61
\times 10^{-2}~\eV
\; , \nonumber\\
m^{}_2 &\approx& m^{}_3 \zeta ~~ \approx ~~ 5.15 \times 10^{-2}~\eV
\; , \nonumber\\
m^{}_1 &\approx& m^{}_3 \xi ~~ \approx ~~ 5.11 \times 10^{-2}~\eV
\;,
%     (65)
\end{eqnarray}
which are nearly degenerate. As a consequence, the effective
neutrino mass in the $0\nu2\beta$ decays $\langle m\rangle^{}_{\rm
ee} \approx m^{}_3 |\tan^2 \theta^{}_{23} e^{i\delta} - 2 \cot
2\theta^{}_{12} \sec \theta^{}_{23}| \approx 2.2 \times 10^{-2}~\eV$
could be probed in the future $0\nu2\beta$ decay experiments.
\end{itemize}

The above analytical analyses serve as an explicit example for how
to determine the leptonic CP-violating phases and neutrino masses,
when a hybrid texture of $M^{}_\nu$ is taken. The full parameter
space of these patterns will be analyzed in the following section.

\section{Numerical Results}

As mentioned before, we have performed a numerical study of all the
sixty hybrid textures of $M^{}_\nu$. It turns out that thirty-nine
of them are consistent with current neutrino oscillation data at the
$3\sigma$ level, while only thirteen patterns can survive if the
$1\sigma$ ranges of neutrino mixing parameters are considered. Our
numerical analysis has been done in the following way:
\begin{enumerate}[1)]
\item For each pattern of $M^{}_\nu$, we generate a set of random
numbers for two neutrino mass-squared differences $(\delta m^2,
\Delta m^2)$ and three neutrino mixing angles $(\theta^{}_{12},
\theta^{}_{23}, \theta^{}_{13})$ in their $3\sigma$ ranges
\cite{Fogli}, which have been shown in Table 1. As we have shown in
the previous section, it is then possible to determine the Dirac
CP-violating phase $\delta$ from Eq.~(12). Instead, we generate a
random number of $\delta$ in the range of $[0, 2\pi)$, and test
whether Eq.~(12) is satisfied by the generated random numbers.

\item In practice, with the random numbers above, we first calculate
the neutrino mass ratios $\zeta$ and $\xi$, and then three neutrino
masses $(m^{}_1, m^{}_2, m^{}_3)$. The criteria for whether a
specific pattern of $M^{}_\nu$ is consistent with current
experimental data are set as follows: (a) With $\zeta$ and $\xi$, we
can determine from Eq.~(12) the value of $R^{}_\nu$, which is
required to fall into the range $\left[(\delta
m^2)^{3\sigma}_\textrm{\fns min} / (\Delta
m^2)^{3\sigma}_\textrm{\fns max}, (\delta
m^2)^{3\sigma}_\textrm{\fns max} / (\Delta
m^2)^{3\sigma}_\textrm{\fns min} \right]$. (b) Since only two
possible neutrino mass hierarchies $m^{}_1 < m^{}_2 < m^{}_3$ and
$m^{}_3 < m^{}_1 < m^{}_2$ are allowed by neutrino oscillation
experiments, we further demand $(\xi^2 - 1) (\zeta^2 - 1) > 0$,
namely, $\xi^2 < \zeta^2 < 1$ corresponds to the normal mass
hierarchy while $\zeta^2 > \xi^2 > 1$ to the inverted mass
hierarchy. (c) The absolute scale of neutrino masses receives
constraints from the beta-decay and $0\nu2\beta$-decay experiments,
however, the most restrictive one comes from cosmological
observations \cite{WMAP9}. Recently, the Planck Collaboration has
released the first data on cosmic microwave background (CMB)
\cite{Planck}. Combined with the WMAP-polarization, high-resolution
CMB, and BAO data sets, the Planck data have placed an upper bound
on the sum of three neutrino masses $\sum m^{}_i < 0.23~\eV$ at the
$95\%$ confidence level. We require that this upper bound should be
satisfied. Once one pattern of $M^{}_\nu$ survives all the above
constraints, we will calculate its predictions for the effective
neutrino mass $\langle m\rangle^{}_\textrm{\fns ee}$ in the
$0\nu2\beta$ decays, the two Majorana CP-violating phases $(\rho,
\sigma)$ and the Jarlskog invariant for leptonic CP violation
$J^{}_\textrm{\fns CP} = s^{}_{12} c^{}_{12} s^{}_{23} c^{}_{23}
s^{}_{13} c^2_{13} \sin\delta$ \cite{Jarlskog}. In addition, the
allowed parameter space of other mixing parameters can be
determined.

\item To illustrate our results, we present a series of figures
for each viable pattern in Figs.~\ref{fig.A1}--\ref{fig.E8}, where
each figure consists of twelve plots in four rows. The first row
shows the allowed ranges of three flavor mixing angles
$\theta^{}_{12}$, $\theta_{23}$ and $\theta_{13}$, versus the Dirac
CP-violating phase $\delta$. In the second row, the histograms of
three mixing angles are given, indicating their distributions in the
allowed parameter space. As the ongoing and forthcoming neutrino
oscillation experiments will provide us with more precise
measurements of three neutrino mixing angles and the Dirac
CP-violating phase, our numerical illustrations make it easy to see
whether a currently viable hybrid texture can be ruled out by future
experimental data. In the third row, we present the allowed ranges
of three neutrino mass eigenvalues $(m^{}_1, m^{}_2, m^{}_3)$ and
the effective neutrino mass $\langle m \rangle^{}_{\rm ee}$ versus
$\delta$. From these plots, one can immediately figure out which
neutrino mass hierarchy is predicted, and whether the effective
neutrino mass $\langle m \rangle^{}_{\rm ee}$ is accessible in
future $0\nu2\beta$ experiments. Finally, the Majorana CP-violating
phases $\rho$, $\sigma$ and the Jarlskog invariant $J^{}_{\rm CP}$
are depicted in the last row. The next-generation long-baseline
neutrino oscillation experiments are promising to discover the
leptonic CP violation if the Jarlskog invariant $J^{}_{\rm CP}$ is
at the percent level \cite{Branco}.
\end{enumerate}

We have carried out a thorough numerical study of all the sixty
patterns of $M^{}_\nu$, however, it will render our paper unreadable
if all the figures of the viable thirty-nine patterns are presented.
Therefore, we will focus only on the six patterns ${\bf A}^{}_1$,
${\bf B}^{}_1$, ${\bf B}^{}_5$, ${\bf D}^{}_1$, ${\bf E}^{}_1$, and
${\bf E}^{}_8$, for which the analytical results have been given in
the previous section. Some comments and discussions on the numerical
results in Figs.~\ref{fig.A1}--\ref{fig.E8} are in order.
\begin{itemize}
\item \textbf{Pattern A$_1$} -- As shown in first row of Fig.~\ref{fig.A1},
the Dirac CP-violating phase $\delta$ is essentially unconstrained
and the whole range $[0, 2\pi)$ of $\delta$ is experimentally
allowed. This can be easily understood from Eq.~(24), where
$R^{}_\nu$ is found to be independent of $\delta$ and
$\theta^{}_{13}$ at the leading order. In contrast with the mixing
angles $\theta^{}_{12}$ and $\theta^{}_{13}$, which are only mildly
constrained, the allowed range of $\theta^{}_{23}$ splits into two
distinct branches: one for $\theta^{}_{23} < 45^\circ$, and the
other for $\theta^{}_{23} < 45^\circ$. Moreover, the histogram of
$\theta^{}_{23}$ in the second row indicates that $\theta^{}_{23} <
45^\circ$ is preferred, in particular the values as small as
$\theta_{23}=35.1^\circ$ at the lower border of the $3\sigma$ range.
Consequently, the precision measurement of $\theta^{}_{23}$ can
finally tell us whether this pattern is allowed or not. It is also
interesting that the histogram of $\theta^{}_{13}$ peaks around
$\theta^{}_{13} = 8^\circ$, quite close to the best-fit value. Three
neutrino masses are given in the third row, where one can observe
that only the normal mass hierarchy (i.e., $m^{}_1 < m^{}_2 <
m^{}_3$) is possible. For {\bf Pattern} ${\bf A}^{}_1$, the
effective neutrino mass in $0\nu2\beta$ decays is exactly zero,
implying some cancellation takes place among the contributions of
three neutrino mass eigenstates. In the fourth row, we can see that
an approximately linear correlation exists between the Majorana CP
phases $(\rho, \sigma)$ and $\delta$. For a maximal CP-violating
phases $\delta = \pi/2$ or $3\pi/2$, the Jarlskog invariant
$|J^{}_{\rm CP}| \sim 3\%$ can be achieved.

\item \textbf{Pattern B$_1$} -- We can see clearly from the first row of
Fig.~\ref{fig.B1} that only two narrow ranges around $\delta =
\pi/2$ and $ \delta = 3\pi/2$ are experimentally allowed. Although
three neutrino mixing angles $\theta^{}_{12}$, $\theta^{}_{23}$ and
$\theta^{}_{13}$ turn out to be arbitrary in their $3\sigma$ ranges,
the distributions of $\theta^{}_{23}$ and $\theta^{}_{13}$ seem to
peak around their best-fit values (i.e., $\theta^{}_{23} =
38.4^\circ$ and $\theta^{}_{13} = 8.9^\circ$), as shown in the
second row. Similar to {\bf Pattern} ${\bf A}^{}_1$, only the normal
neutrino mass hierarchy (i.e., $m^{}_1 < m^{}_2 < m^{}_3$) is
possible, so the effective neutrino mass $\langle m \rangle^{}_{\rm
ee}$ is in general small, which renders it very challenging to
observe the $0\nu2\beta$ decays. Since the constraint on $\delta$ is
quite restrictive, only a small faction of the parameter space of
$\rho$ and $\sigma$ is allowed. Interestingly, the Jarlskog
invariant $J_\textrm{\fns CP}$ is predicted to be close to its
maximum, i.e., $|J^{}_{\rm CP}| \geq 3\%$, which should be
accessible to the next-generation long-baseline neutrino oscillation
experiments \cite{Branco}.

\item \textbf{Pattern B$_5$} -- From the first row of Fig.~\ref{fig.B6},
one can observe that the allowed ranges of $\delta$ contain two
disjointed regions: one around $\delta = \pi$, and the other around
$\delta = 0$ or $2\pi$. This feature becomes clear, if we recall the
analytical discussions on {\bf Pattern} ${\bf B}^{}_5$ in the
previous section, where $\cos\delta > 0$ (i.e., $\delta < \pi/2$ or
$\delta>3\pi/2$) for $\theta^{}_{23} < 45^\circ$, and $\cos\delta <
0$ (i.e., $\pi/2 < \delta < 3\pi/2$) for $\theta^{}_{23} >
45^\circ$, are required to guarantee $m^{}_2 > m^{}_1$. While both
$\theta^{}_{12}$ and $\theta^{}_{13}$ are mildly constrained,
$\theta^{}_{23}$ is restricted to smaller regions far from the
maximal mixing. Moreover, there is a strong correlation between
$\theta^{}_{23}$ and $\delta$. For instance, if $\delta \approx \pi$
is taken, then $\theta^{}_{23} \approx 51^\circ$ holds; if $\delta
\approx 0$ or $2\pi$ is assumed, then we obtain $\theta^{}_{23}
\approx 39^\circ$. As shown in the second row, the distribution of
$\theta^{}_{23}$ peaks around its best-fit value, namely,
$\theta^{}_{23} = 38.4^\circ$. The numerical results in the third
row indicate that only the normal mass hierarchy ($m^{}_1 < m^{}_2 <
m^{}_3$) is allowed, and the effective neutrino mass $\langle m
\rangle^{}_{\rm ee}$ is quite small. The Majorana CP-violating
phases $(\rho, \sigma)$ and the Jarlskog invariant $J^{}_{\rm CP}$
are shown in the last row, where one can observe that the parameter
space is tightly constrained.

\item \textbf{Pattern D$_1$} -- The Dirac CP-violating phase $\delta$
falls into $(\pi/2, 3\pi/2)$, as shown in the first row of
Fig.~\ref{fig.D1}. It is evident from Eq.~(49) that only $\cos
\delta <0$ (i.e., $\pi/2 < \delta < 3\pi/2$) is possible, given the
$3\sigma$ ranges of mixing parameters. Although the $3\sigma$ ranges
of three mixing angles are still allowed, only smaller (larger)
values of $\theta^{}_{12}$ ($\theta^{}_{23}$) can survive if $\delta
\approx \pi$ is assumed. In the second row, the distributions of
three mixing angles do not show significant preference in any
specific ranges. Different from the previous patterns of $M^{}_\nu$,
{\bf Pattern} ${\bf D}^{}_1$ predicts the inverted neutrino mass
hierarchy ($m^{}_3 < m^{}_1 < m^{}_2$), as indicated in the third
row. Consequently, the effective neutrino mass $\langle m
\rangle^{}_{\rm ee}$ is larger than $0.02~\eV$, which is reachable
in next-generation $0\nu2\beta$ decay experiments. While $\rho$ is
restricted to a relatively small range around $\pi/2$ or $-\pi/2$,
$\sigma$ can take any values in $[-\pi/4, \pi/4]$. In addition, the
magnitude of Jarlskog invariant $J^{}_{\rm CP}$ could be at the
percent level if $\delta$ is close to $\pi/2$ or $3\pi/2$.

\item \textbf{Pattern E$_1$} -- As shown in the first row of
Fig.~\ref{fig.E1}, the Dirac CP-violating phase $\delta$ is limited
to two small regions around $3\pi/4$ and $5\pi/4$. From the second
row, we can observe that $\theta^{}_{12}$ and $\theta^{}_{13}$ show
no significant preference in their $3\sigma$ ranges, while
$\theta^{}_{23}$ has two peaks around $37^\circ$ and $51^\circ$.
Note that the former peak dominates over the latter and is quite
close to the best-fit value of $\theta^{}_{23}$. The neutrino mass
spectrum is nearly degenerate, namely, $m^{}_1 \approx m^{}_2
\approx m^{}_3 \approx 0.1~\eV$, as shown in the third row.
Therefore, the effective neutrino mass $\langle m \rangle^{}_{\rm
ee}$ can be as large as $0.05~\eV$, which is much larger than those
in all the previous patterns. In addition, the parameter space of
two Majorana CP-violating phases $(\rho, \sigma)$ and the Jarlskog
invariant $J^{}_{\rm CP}$ is strongly constrained. Since $\sin
\delta$ is nonzero, the leptonic CP violation is expected.

\item \textbf{Pattern E$_8$} -- From the first row of Fig.~\ref{fig.E8},
one can see that the Dirac CP-violating phase $\delta$ is restricted
to the ranges of $\delta < \pi/2$ and $\delta > 3\pi/2$, which are
consistent with our analytical results in section 3.  The
distribution of $\theta^{}_{23}$ peaks around $37^\circ$, while
those of $\theta^{}_{12}$ and $\theta^{}_{13}$ do not show
significant preference, as shown in the second row. Only the
inverted neutrino mass hierarchy ($m^{}_3 < m^{}_1 < m^{}_2$) is
allowed, and the effective neutrino mass $\langle m \rangle^{}_{\rm
ee}$ turns out to be in the range $[0.025~\eV, 0.040~\eV]$, which is
accessible in future $0\nu2\beta$ decay experiments. In the last
row, the two Majorana CP-violating phases $\rho$ and $\sigma$ are
found to be linearly correlated with the Dirac CP phase $\delta$.
\end{itemize}

It is worthwhile to stress that the precision measurements of three
neutrino mixing angles, in particular the octant of
$\theta^{}_{23}$, as well as the discovery of leptonic CP violation
and the $0\nu2\beta$ decays, are crucially important to distinguish
between different hybrid textures of Majorana neutrino mass matrix.
For instance, we have also checked that only thirteen hybrid
textures (i.e., ${\bf A}^{}_3$, ${\bf C}^{}_{2,5}$, ${\bf
D}^{}_{1,5,6}$, ${\bf E}^{}_{1,4,6,7}$, and ${\bf F}^{}_{1,7,8}$)
are viable, if the $1\sigma$ ranges of neutrino mixing parameters
and the Dirac CP-violating phase are taken into account
\cite{Fogli}. In addition, the cosmological bound on absolute
neutrino masses becomes very relevant. If the upper limit $\sum
m^{}_i < 0.44~\eV$ from the nine-year WMAP observations \cite{WMAP9}
is used instead of the latest Planck result, two extra patterns
\begin{eqnarray}
{\bf B}^{}_6: \left(\matrix{
    \triangle & 0 & \times \cr
    0 & \times & \triangle \cr
    \times & \triangle & \times}
  \right) \; , ~~~~ {\bf C}^{}_6:\left(\matrix{
    \triangle & \times & 0 \cr
    \times & \times & \triangle \cr
    0 & \triangle & \times}
  \right) \; ,
%     (66)
\end{eqnarray}
can survive the oscillation data at the $3\sigma$ level.

\section{Flavor Symmetry}

It has been proved that the texture zeros in the Majorana neutrino
mass matrix $M^{}_\nu$ can be realized by implementing the $Z^{}_n$
flavor symmetry \cite{Grimus1,Grimus2}. Taking the two-zero textures
for example, one can demonstrate that the seven viable patterns can
be derived from $Z^{}_n$ symmetries in the type-II seesaw model
\cite{FXZ}. To illustrate how to realize the hybrid textures of
$M^{}_\nu$, we will work in the type-II seesaw model, which extends
the scalar sector of the standard model with one or more
$SU(2)^{}_{\rm L}$ scalar triplets \cite{SS2}. For $N$ scalar
triplets, the gauge-invariant Lagrangian relevant for neutrino
masses reads
\begin{equation}
-{\cal L}^{}_{\Delta} = \frac{1}{2} \sum_{j} \sum_{\alpha, \beta}
\left(Y^{}_{\Delta^{}_j}\right)^{}_{\alpha \beta}
\overline{\ell^{}_{\alpha \rm L}} \Delta^{}_j i\sigma^{}_2
\ell^c_{\beta \rm L} + {\rm h.c.} \; ,
%     (67)
\end{equation}
where $\alpha$ and $\beta$ run over $e$, $\mu$ and $\tau$,
$\Delta^{}_j$ denotes the $j$-th triplet scalar field (for $j = 1,
2, \cdots, N$), and $Y^{}_{\Delta^{}_j}$ is the corresponding Yukawa
coupling matrix. After the triplet scalar acquires its vacuum
expectation value $\langle \Delta^{}_j\rangle \equiv v^{}_j$, the
Majorana neutrino mass matrix is given by
\begin{equation}
M^{}_\nu = \sum_j Y^{}_{\Delta^{}_j} v^{}_j \; ,
%     (68)
\end{equation}
where the smallness of $v^{}_j$ is attributed to the largeness of
the triplet scalar mass scale \cite{SS2}.

It is worth mentioning that an equality between two matrix elements
cannot be achieved by imposing any Abelian symmetries on the
Lagrangian in Eq.~(67), since the Abelian symmetry group has only
one-dimensional irreducible representations and the Yukawa couplings
for different representations are not necessarily the same. In this
work, we just take the pattern ${\bf A}^{}_2$ as a typical example,
i.e.,
\begin{equation}
M^{{\bf A}^{}_2}_\nu = \left(\matrix{ 0 & a & b \cr a & d & c \cr b
& c & d }\right) \;.
%     (69)
\end{equation}
In order to ensure $(M^{}_\nu)^{}_{ee} = 0$, we have to implement a
$Z^{}_3$ symmetry. Furthermore, to guarantee $(M^{}_\nu)^{}_{\mu
\mu} = (M^{}_\nu)^{}_{\tau \tau}$, we can make use of an $S^{}_3$
symmetry, which is the simplest discrete non-Abelian group. In the
framework of type-II seesaw model, we find that at least five scalar
triplets $\Delta^{}_i$ (for $i=1,2,3,4,5$) are necessary. The
assignments of the scalar triplets and lepton doublets under the
$Z^{}_3$ symmetry are given as follows:
\begin{eqnarray}
\ell^{}_{e {\rm L}} \sim \omega \; , ~~~~~ \ell^{}_{\mu {\rm L}},~
\ell^{}_{\tau {\rm L}} \sim \omega^2 \; , ~~~~~ \Delta^{}_1,~
\Delta^{}_2 \sim 1 \; , ~~~~~ \Delta^{}_3, ~\Delta^{}_4,
~\Delta^{}_5 \sim \omega \; ;
%     (70)
\end{eqnarray}
while the assignments under the $S^{}_3$ symmetry are
\begin{eqnarray}
\ell^{}_{e {\rm L}} \sim {\bf 1} \; , ~~~~~ \left(\matrix{
\ell^{}_{\mu {\rm L}} \cr ~ \cr \ell^{}_{\tau {\rm L}}} \right) \sim
{\bf 2} \; , ~~~~~ \left(\matrix{ \Delta^{}_1 \cr ~ \cr \Delta^{}_2}
\right) \sim {\bf 2} \; , ~~~~~ \left(\matrix{ \Delta^{}_3 \cr ~ \cr
\Delta^{}_4} \right) \sim {\bf 2} \; , ~~~~~ \Delta^{}_5 \sim {\bf
1} \; .
%     (71)
\end{eqnarray}
Hence the $S^{}_3 \otimes Z^{}_3$-invariant Lagrangian relevant for
neutrino masses reads
\begin{eqnarray}
{\cal L}^{}_\nu &=& -\frac{1}{2} y^{}_1 \overline{\ell^{}_{e {\rm
L}}} \left( \Delta^{}_1 \ell^c_{\mu {\rm L}} + \Delta^{}_2
\ell^c_{\tau {\rm L}} \right) -\frac{1}{2} y^{}_2
\left(\overline{\ell^{}_{\mu {\rm L}}} \Delta^{}_5 \ell^c_{\mu {\rm
L}} + \overline{\ell^{}_{\tau {\rm L}}} \Delta^{}_5 \ell^c_{\tau
{\rm L}}\right) \nonumber \\
&~& -\frac{1}{2} y^{}_3 \left[\left(\overline{\ell^{}_{\mu {\rm L}}}
\Delta^{}_3 \ell^c_{\tau {\rm L}} + \overline{\ell^{}_{\tau {\rm
L}}} \Delta^{}_3 \ell^c_{\mu {\rm L}}\right) +
\left(\overline{\ell^{}_{\mu {\rm L}}} \Delta^{}_4 \ell^c_{\mu {\rm
L}} - \overline{\ell^{}_{\tau {\rm L}}} \Delta^{}_4 \ell^c_{\tau
{\rm L}}\right)\right] + {\rm h.c.} \; ,
%     (72)
\end{eqnarray}
where we have used the tensor product ${\bf 2} \otimes {\bf 2} =
{\bf 1} + {\bf 1}^\prime + {\bf 2}$ for the $S^{}_3$ symmetry group
\cite{flavorsym}. After the triplet scalar acquires its vacuum
expectation value $\langle \Delta^{}_i \rangle = v^{}_i$, the
Majorana neutrino mass matrix is given by
\begin{equation}
M^{}_\nu = \left(\matrix{ 0 & y^{}_1 v^{}_1 & y^{}_1 v^{}_2 \cr
y^{}_1 v^{}_1 & y^{}_2 v^{}_5 + y^{}_3 v^{}_4 & 2y^{}_3 v^{}_3 \cr
y^{}_1 v^{}_2 & 2y^{}_3 v^{}_3 & y^{}_2 v^{}_5 - y^{}_3
v^{}_4}\right) \; .
%     (73)
\end{equation}
Once $v^{}_4 = 0$ is obtained by minimizing the scalar potential, we
arrive at the pattern ${\bf A}^{}_2$ in Eq.~(69), with $a = y^{}_1
v^{}_1$, $b = y^{}_1 v^{}_2$, $c = 2y^{}_3 v^{}_3$, and $d = y^{}_2
v^{}_5$. On the other hand, we have to ensure that the
charged-lepton mass matrix $M^{}_l$ is diagonal. This can be
achieved if one introduces two $SU(2)^{}_{\rm L}$ scalar doublets
$\Phi^{}_1$ and $\Phi^{}_2$, in addition to the standard-model Higgs
doublet $H$, which is a singlet under the flavor symmetry. These two
extra scalar doublets and three right-handed charged-lepton fields
are assigned as:
\begin{equation}
e^{}_{\rm R} \sim \omega \; , ~~~ \mu^{}_{\rm R} \;, \tau^{}_{\rm R}
\sim \omega^2 \; , ~~~ \Phi^{}_1, \Phi^{}_2 \sim 1 \;
%     (74)
\end{equation}
under the $Z^{}_3$ symmetry, while
\begin{equation}
e^{}_{\rm R}, \tau^{}_{\rm R} \sim {\bf 1} \; ,  ~~~ \mu^{}_{\rm R}
\sim {\bf 1}^\prime \; , ~~~ \left(\matrix{ \Phi^{}_1 \cr ~ \cr
\Phi^{}_2} \right) \sim {\bf 2} \;
%     (75)
\end{equation}
under the $S^{}_3$ symmetry. Three lepton doublets transform in the
same way as in Eqs.~(70) and (71). Therefore, the $S^{}_3 \otimes
Z^{}_3$-invariant Lagrangian relevant for the charged-lepton masses
is
\begin{equation}
{\cal L}^{}_l = - y^{}_e \overline{\ell^{}_{e{\rm L}}} H e^{}_{\rm
R} - y^{}_\mu \left(\overline{\ell^{}_{\mu {\rm L}}} \Phi^{}_2 -
\overline{\ell^{}_{\tau {\rm L}}} \Phi^{}_1\right) \mu^{}_{\rm R} -
y^{}_\tau \left(\overline{\ell^{}_{\mu {\rm L}}} \Phi^{}_1 +
\overline{\ell^{}_{\tau {\rm L}}} \Phi^{}_2\right) \tau^{}_{\rm R} +
{\rm h.c.} \; ,
%     (76)
\end{equation}
leading to a diagonal charged-lepton mass matrix $M^{}_l \equiv {\rm
Diag}\{m^{}_e, m^{}_\mu, m^{}_\tau\}$ with three mass eigenvalues
$m^{}_e = y^{}_e v/\sqrt{2}$, $m^{}_\mu = y^{}_\mu u/\sqrt{2}$, and
$m^{}_\tau = y^{}_\tau u /\sqrt{2}$, where the vacuum expectation
values of three scalar doublets $\langle H \rangle = v/\sqrt{2}$,
$\langle \Phi^{}_2 \rangle = u/\sqrt{2}$ and $\langle \Phi^{}_1
\rangle = 0$ have been taken. Our numerical analysis has
demonstrated that the hybrid texture in Eq.~(69) is consistent with
current neutrino oscillation data at the $3\sigma$ level.

To complete the above $S^{}_3 \otimes Z^{}_3$ flavor model, we have
to examine in detail the invariant scalar potential and check
whether vacuum alignments $v^{}_1 \neq v^{}_2$, $v^{}_3 \neq
v^{}_4$, and $\langle \Phi^{}_1 \rangle \neq \langle \Phi^{}_2
\rangle$ can be actually realized. For this purpose, one can work in
the supersymmetric version of type-II seesaw model and follow the
method of driving flavon fields, as proposed in
Refs.~\cite{Altarelli,flavorsym1}. In addition, the symmetry
realization of all the thirty-nine viable hybrid textures of
$M^{}_\nu$ in a systematic way deserves further studies and will be
discussed elsewhere.

\section{Summary}

\begin{figure}[]
\centering
\includegraphics[width=0.9\textwidth]{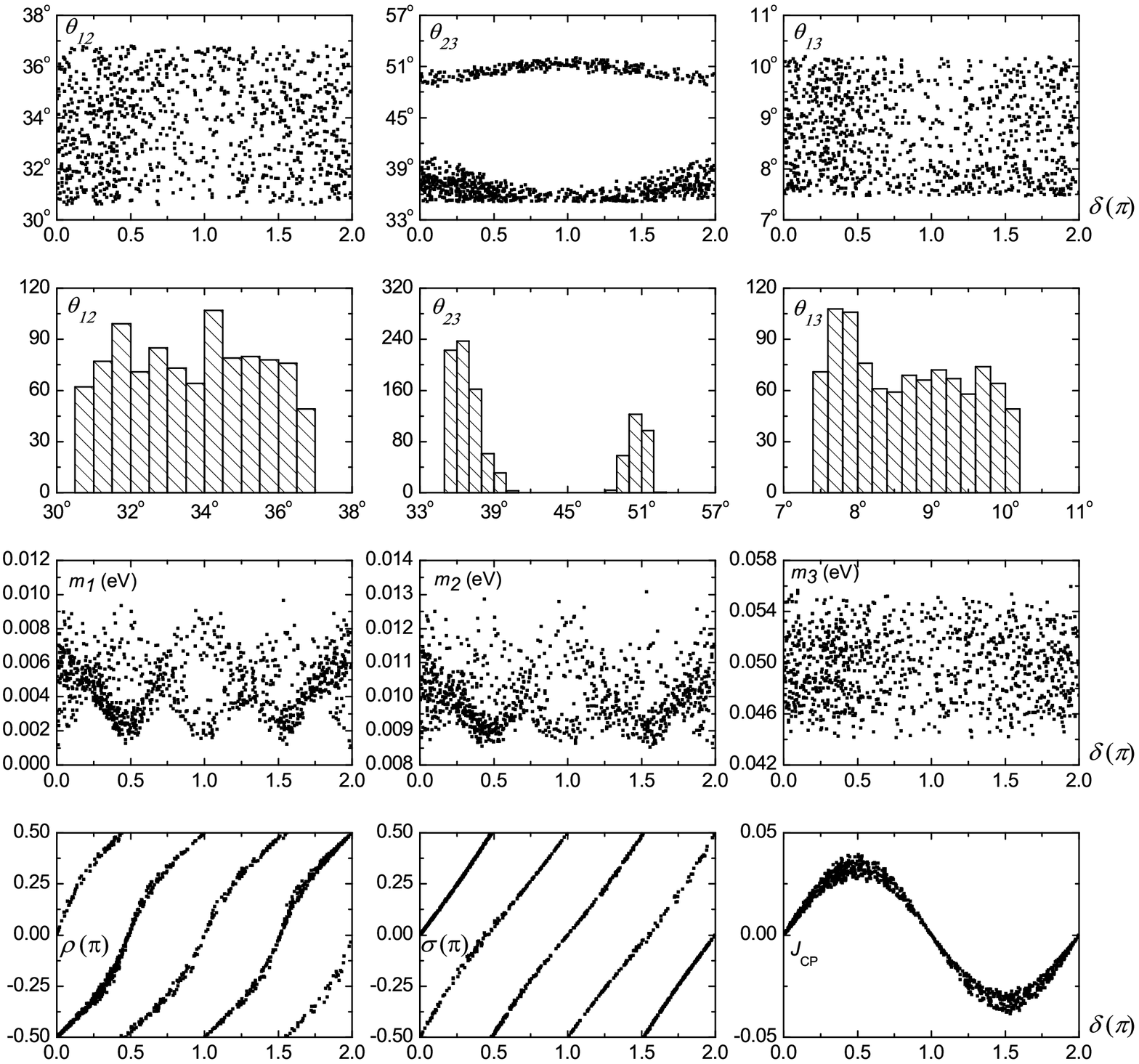}
\caption{Pattern ${\bf A}^{}_1$ of $M^{}_\nu$: The allowed ranges of
flavor mixing angles $(\theta^{}_{12}, \theta^{}_{23},
\theta^{}_{13})$ versus the Dirac CP-violating phase $\delta$ at the
$3\sigma$ level, and the probability distribution of three angles,
are given in the first and second rows, respectively. In the third
and fourth rows, the predictions for three neutrino masses $(m^{}_1,
m^{}_2, m^{}_3)$, the Majorana CP-violating phases $(\rho, \sigma)$
and the Jarlskog invariant $J^{}_{\rm CP}$ are shown with respect to
the Dirac CP-violating phase $\delta$.} \label{fig.A1}
\end{figure}

\begin{figure}[]
\centering
\includegraphics[width=0.9\textwidth]{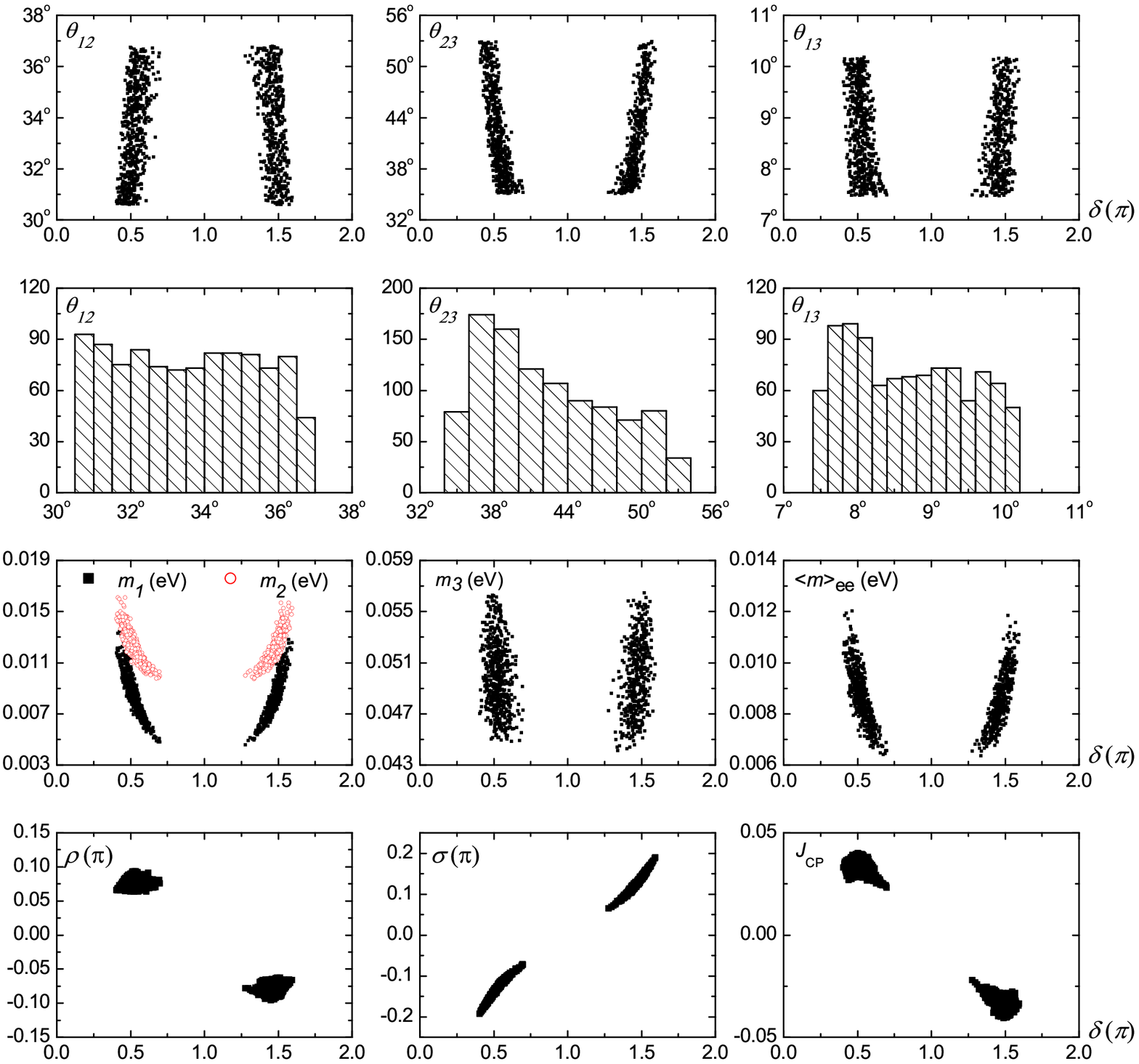}
\caption{Pattern ${\bf B}^{}_1$ of $M^{}_\nu$: The allowed ranges of
flavor mixing angles $(\theta^{}_{12}, \theta^{}_{23},
\theta^{}_{13})$ versus the Dirac CP-violating phase $\delta$ at the
$3\sigma$ level, and the probability distribution of three angles,
are given in the first and second rows, respectively. In the third
and fourth rows, the predictions for three neutrino masses $(m^{}_1,
m^{}_2, m^{}_3)$ and the effective neutrino mass in neutrinoless
double-beta decays $\langle m \rangle^{}_{\rm ee}$, the Majorana
CP-violating phases $(\rho, \sigma)$ and the Jarlskog invariant
$J^{}_{\rm CP}$, are shown with respect to the Dirac CP-violating
phase $\delta$.} \label{fig.B1}
\end{figure}

\begin{figure}[]
\centering
\includegraphics[width=0.9\textwidth]{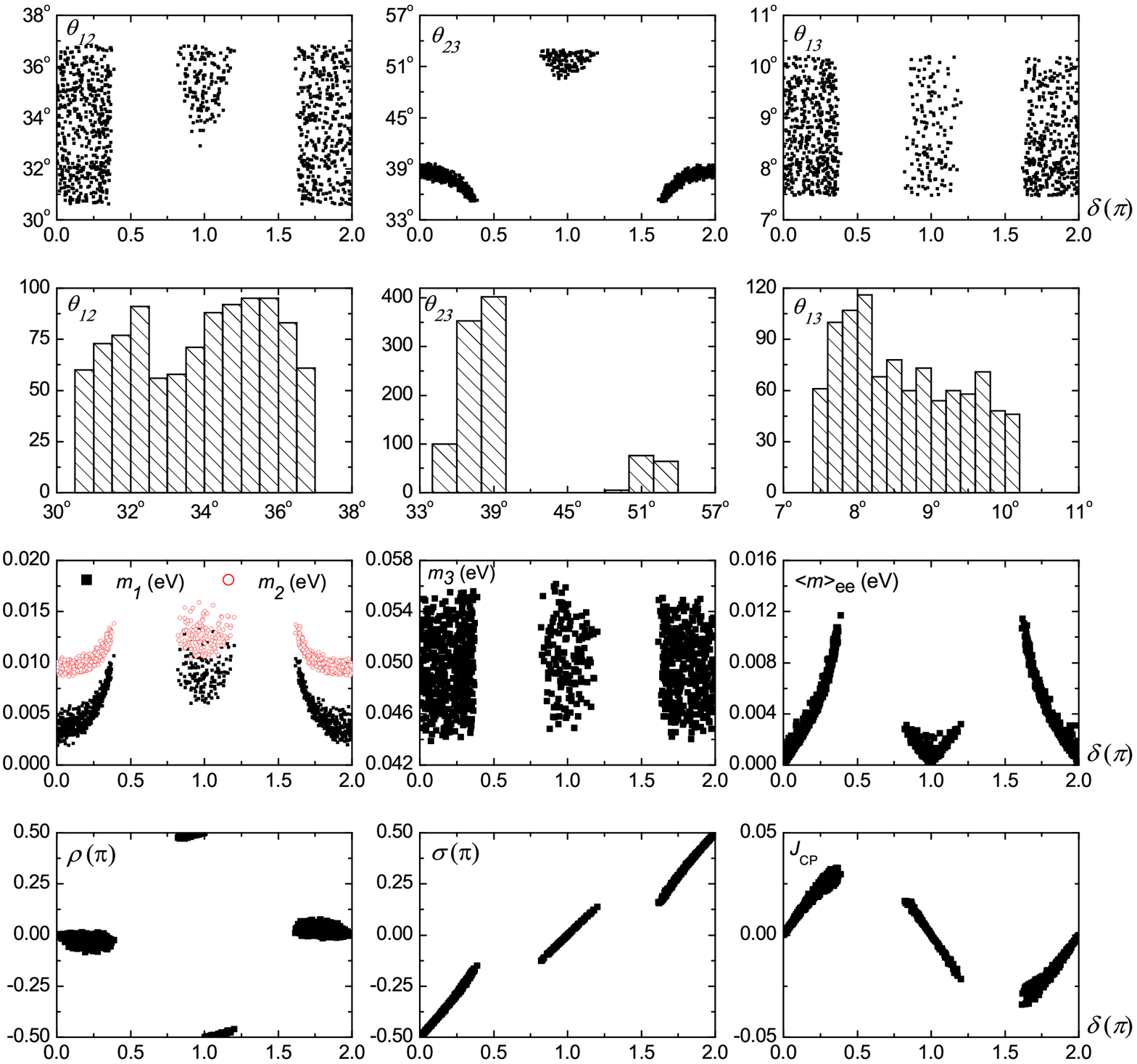}
\caption{Pattern ${\bf B}^{}_5$ of $M^{}_\nu$: The allowed ranges of
flavor mixing angles $(\theta^{}_{12}, \theta^{}_{23},
\theta^{}_{13})$ versus the Dirac CP-violating phase $\delta$ at the
$3\sigma$ level, and the probability distribution of three angles,
are given in the first and second rows, respectively. In the third
and fourth rows, the predictions for three neutrino masses $(m^{}_1,
m^{}_2, m^{}_3)$ and the effective neutrino mass in the neutrinoless
double-beta decays $\langle m \rangle^{}_{\rm ee}$, the Majorana
CP-violating phases $(\rho, \sigma)$ and the Jarlskog invariant
$J^{}_{\rm CP}$, are shown with respect to the Dirac CP-violating
phase $\delta$.} \label{fig.B6}
\end{figure}

\begin{figure}[]
\centering
\includegraphics[width=0.9\textwidth]{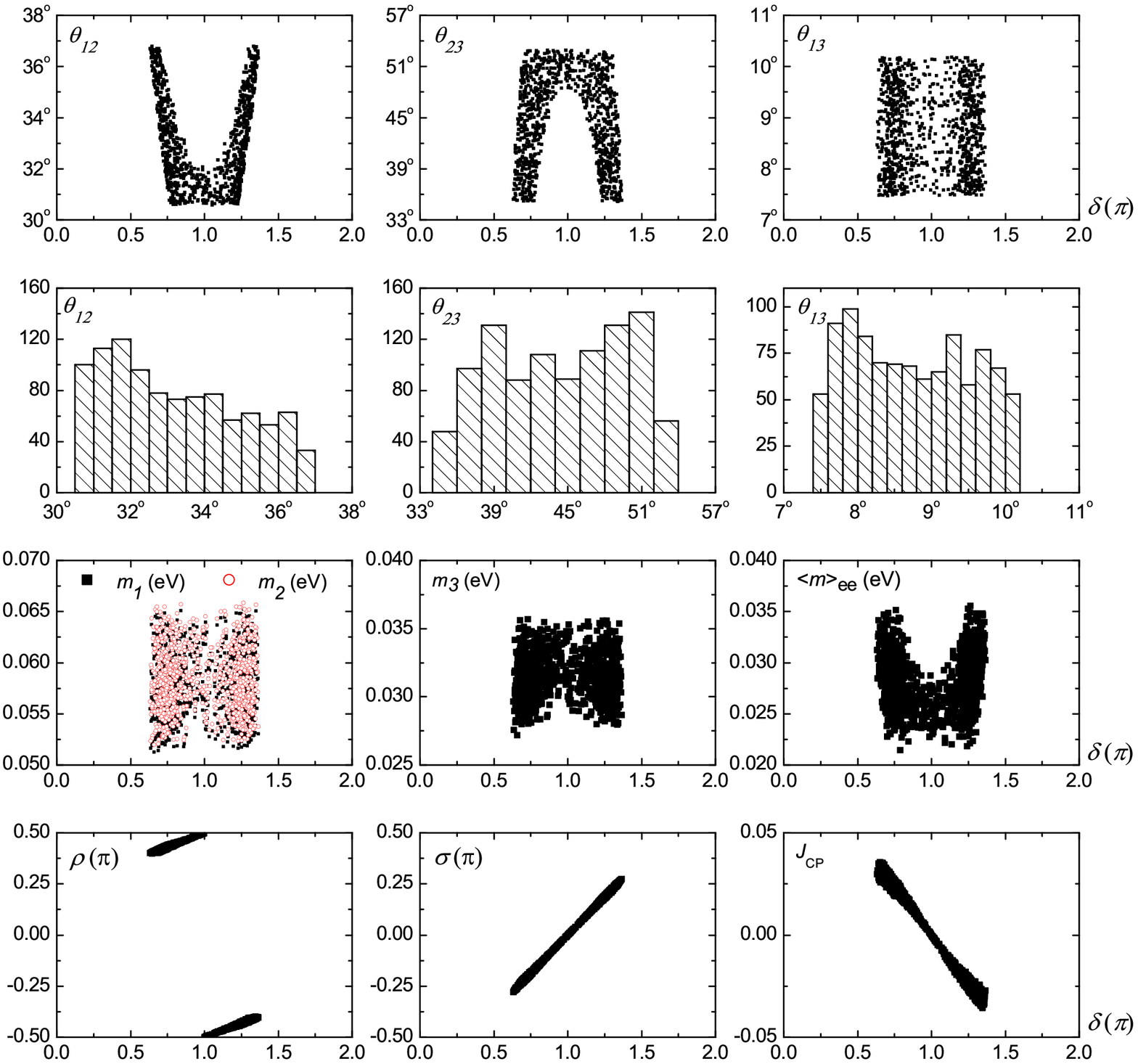}
\caption{Pattern ${\bf D}^{}_1$ of $M^{}_\nu$: The allowed ranges of
flavor mixing angles $(\theta^{}_{12}, \theta^{}_{23},
\theta^{}_{13})$ versus the Dirac CP-violating phase $\delta$ at the
$3\sigma$ level, and the probability distribution of three angles,
are given in the first and second rows, respectively. In the third
and fourth rows, the predictions for three neutrino masses $(m^{}_1,
m^{}_2, m^{}_3)$ and the effective neutrino mass in the neutrinoless
double-beta decays $\langle m \rangle^{}_{\rm ee}$, the Majorana
CP-violating phases $(\rho, \sigma)$ and the Jarlskog invariant
$J^{}_{\rm CP}$, are shown with respect to the Dirac CP-violating
phase $\delta$.} \label{fig.D1}
\end{figure}

\begin{figure}[]
\centering
\includegraphics[width=0.9\textwidth]{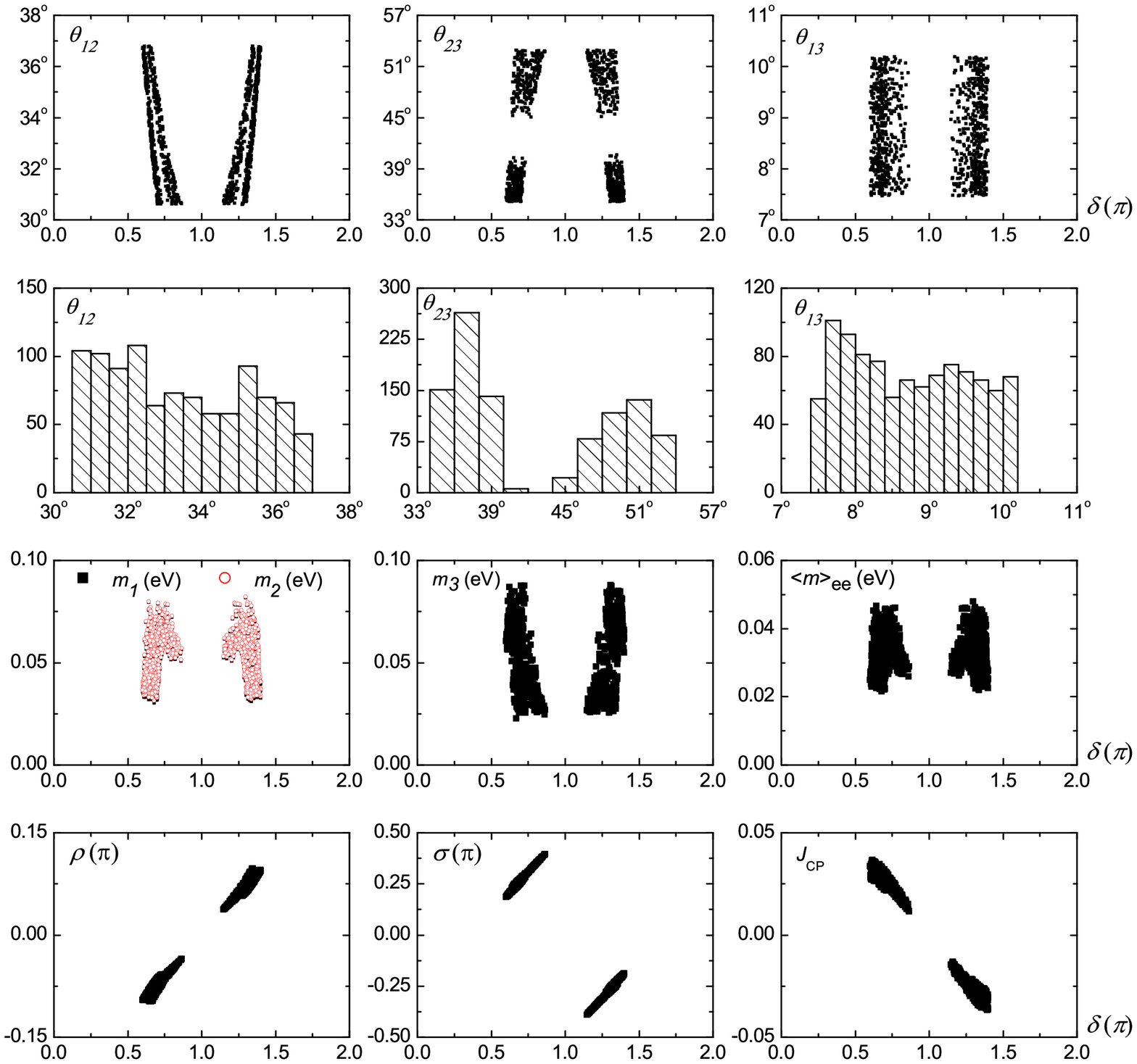}
\caption{Pattern ${\bf E}^{}_1$ of $M^{}_\nu$: The allowed ranges of
flavor mixing angles $(\theta^{}_{12}, \theta^{}_{23},
\theta^{}_{13})$ versus the Dirac CP-violating phase $\delta$ at the
$3\sigma$ level, and the probability distribution of three angles,
are given in the first and second rows, respectively. In the third
and fourth rows, the predictions for three neutrino masses $(m^{}_1,
m^{}_2, m^{}_3)$ and the effective neutrino mass in the neutrinoless
double-beta decays $\langle m \rangle^{}_{\rm ee}$, the Majorana
CP-violating phases $(\rho, \sigma)$ and the Jarlskog invariant
$J^{}_{\rm CP}$, are shown with respect to the Dirac CP-violating
phase $\delta$.} \label{fig.E1}
\end{figure}

\begin{figure}[]
\centering
\includegraphics[width=0.9\textwidth]{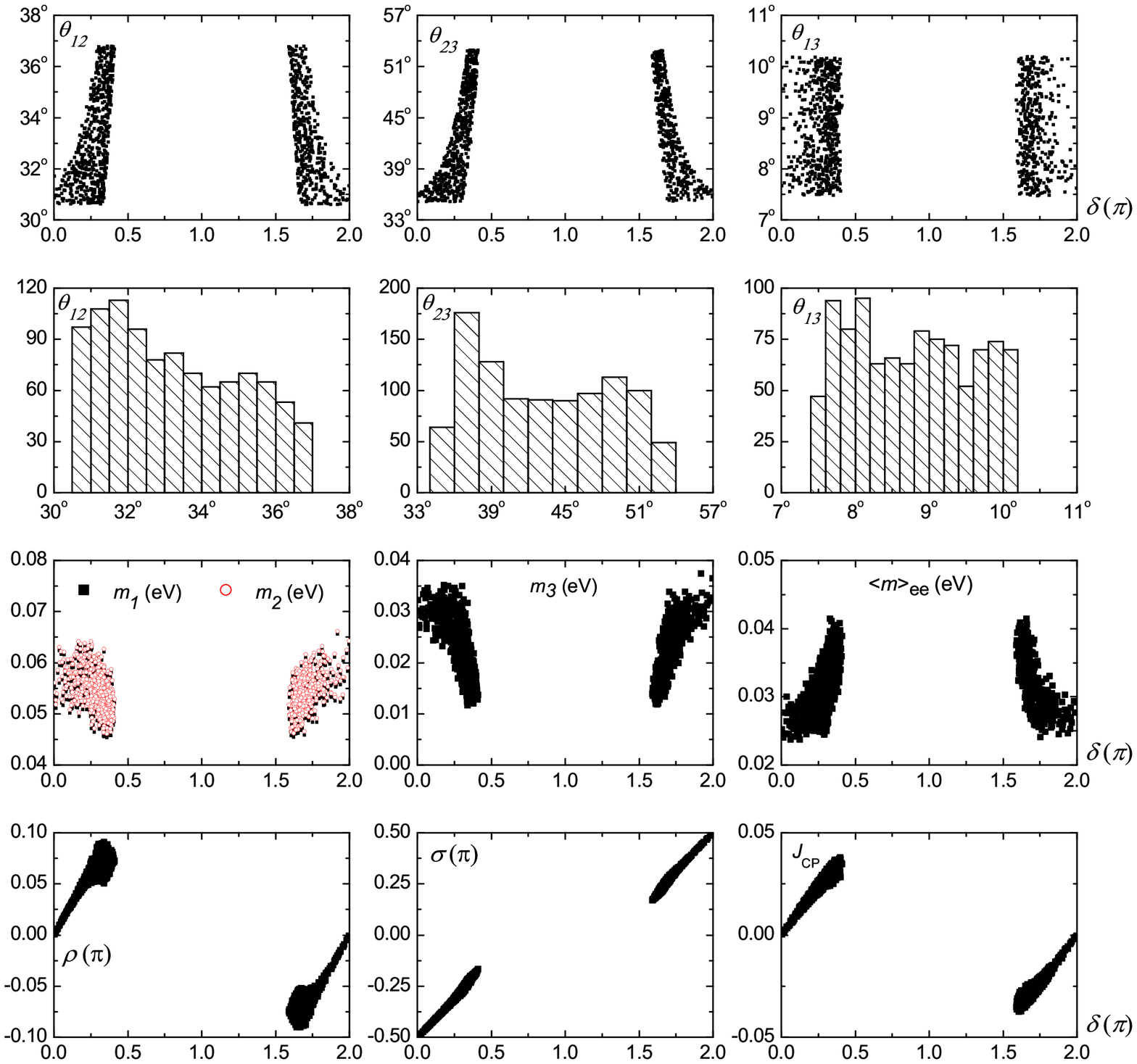}
\caption{Pattern ${\bf E}^{}_8$ of $M^{}_\nu$: The allowed ranges of
flavor mixing angles $(\theta^{}_{12}, \theta^{}_{23},
\theta^{}_{13})$ versus the Dirac CP-violating phase $\delta$ at the
$3\sigma$ level, and the probability distribution of three angles,
are given in the first and second rows, respectively. In the third
and fourth rows, the predictions for three neutrino masses $(m^{}_1,
m^{}_2, m^{}_3)$ and the effective neutrino mass in the neutrinoless
double-beta decays $\langle m \rangle^{}_{\rm ee}$, the Majorana
CP-violating phases $(\rho, \sigma)$ and the Jarlskog invariant
$J^{}_{\rm CP}$, are shown with respect to the Dirac CP-violating
phase $\delta$.} \label{fig.E8}
\end{figure}

Motivated by the recent measurements of $\theta^{}_{13}$ in reactor
neutrino experiments, we have performed a thorough study of the
so-called hybrid textures of Majorana neutrino mass matrix
$M^{}_\nu$, where one texture zero and one equality between two
nonzero matrix elements are assumed. We have found that thirty-nine
out of sixty possible patterns are compatible with current
experimental data at the $3\sigma$ level. In the following, our main
results are summarized:
\begin{itemize}
\item If three neutrino mixing angles $(\theta^{}_{12}, \theta^{}_{23},
\theta^{}_{13})$ and two neutrino mass-squared differences $(\delta
m^2, \Delta m^2)$ are given, the leptonic CP-violating phases
$(\rho, \sigma, \delta)$ and three neutrino masses $(m^{}_1, m^{}_2,
m^{}_3)$ can be fully determined. For illustration, we have chosen
six hybrid textures ${\bf A}^{}_1$, ${\bf B}^{}_1$, ${\bf B}^{}_5$,
${\bf D}^{}_1$, ${\bf E}^{}_1$, and ${\bf E}^{}_8$, and derived the
analytical approximate formulas for the CP-violating phases and
neutrino masses.

\item In Figs.~\ref{fig.A1}--\ref{fig.E8}, we show the numerical
results for the above six hybrid textures. Some interesting
observations should be mentioned: (1) The allowed regions of
$\theta^{}_{23}$ from the patterns ${\bf A}^{}_1$ and ${\bf B}^{}_5$
turn out to be well separated and deviate significantly from the
maximal mixing. Therefore, if $\theta^{}_{23}$ is finally measured
to be very close to $\pi/4$, then both ${\bf A}^{}_1$ and ${\bf
B}^{}_5$ can be excluded. (2) Except the pattern ${\bf A}^{}_1$, all
the hybrid textures under consideration have specific predictions
for the CP-violating phase $\delta$. In addition, their predictions
for the effective neutrino mass $\langle m \rangle^{}_{\rm ee}$
could also be quite different. (3) Now the cosmological bound on the
absolute scale of neutrino masses becomes quite relevant for the
study of lepton flavor structure.
\end{itemize}

Moreover, we have considered the stability of texture zeros and
equality against one-loop quantum corrections. In a type-II seesaw
model, we illustrate how to realize the hybrid texture ${\bf
A}^{}_2$ by implementing an $S^{}_3 \otimes Z^{}_3$ flavor symmetry.

The ongoing and upcoming neutrino oscillation experiments are
expected to precisely measure the neutrino mixing parameters, in
particular the smallest mixing angle $\theta^{}_{13}$, the deviation
of $\theta^{}_{23}$ from $\pi/4$ and the Dirac CP-violating phase
$\delta$. The sensitivity of future cosmological observations to the
sum of neutrino masses $\sum m^{}_i$ and the sensitivity of the
neutrinoless double-beta decay experiments to the effective mass
term $\langle m \rangle^{}_{\rm ee}$ will probably reach
$\sim0.05~{\rm eV}$ in the near future. We therefore expect that
many patterns of the hybrid textures of $M^{}_\nu$ can be excluded
or only marginally allowed by tomorrow's data, and those capable of
surviving should shed light on the underlying flavor structures of
massive neutrinos.

\vspace{0.5cm}
\begin{flushleft}
{\large \bf Acknowledgements}
\end{flushleft}

One of the authors (J.Y.L.) would like to thank Prof. Zhi-zhong Xing
for inspiring discussions, and Theoretical Physics Division of IHEP
for financial support and hospitality during his visit in Beijing,
where part of this work was done. This work was supported in part by
the National Natural Science Foundation of China under grant No.
11205113 (J.Y.L.) and by the G\"{o}ran Gustafsson Foundation (S.Z.).

\newpage

\end{document}